\documentclass[aps,prl,reprint,superscriptaddress]{revtex4-2}
\usepackage{color,hyperref}
\usepackage{graphicx}
\usepackage{amsmath}
\usepackage{amssymb}
\usepackage{mathrsfs}
\usepackage{xcolor}
\usepackage{mathtools}
\usepackage{braket,bm}
\usepackage{longtable}

\begin{document}


\title{High density loading and collisional loss of laser cooled molecules in an optical trap}
%


\author{Varun Jorapur}
\affiliation{Department of Physics, Yale University, New Haven, Connecticut 06511, USA}
\affiliation{Department of Physics, University of Chicago, Chicago, Illinois 60637, USA}
\author{Thomas K. Langin}
\affiliation{Department of Physics, University of Chicago, Chicago, Illinois 60637, USA}
\author{Qian Wang}
\affiliation{Department of Physics, University of Chicago, Chicago, Illinois 60637, USA}
\author{Geoffrey Zheng}
\affiliation{Department of Physics, University of Chicago, Chicago, Illinois 60637, USA}
\author{David DeMille}
\affiliation{Department of Physics, University of Chicago, Chicago, Illinois 60637, USA}

\date{\today}

\begin{abstract}

We report optical trapping of laser-cooled molecules at sufficient density to observe molecule-molecule collisions for the first time in a bulk gas. SrF molecules from a red-detuned magneto-optical trap (MOT) are compressed and cooled in a blue-detuned MOT. Roughly 30\% of these molecules are loaded into an optical dipole trap with peak number density $n_0 \approx 3\times 10^{10} \text{ cm}^{-3}$ and temperature $T\approx40$ $\mu$K. We observe two-body loss with rate coefficient $\beta = 2.7^{+1.2}_{-0.8}\times 10^{-10} \text{ cm}^3 \text{ s}^{-1}$. Achieving this density and temperature opens a path to evaporative cooling towards quantum degeneracy of laser-cooled molecules.
\end{abstract}

\maketitle

Ultracold polar molecules, with their long-range dipolar interactions and rich internal structure, have emerged as a powerful platform for quantum information science, quantum simulation, and precision probes of fundamental physics \cite{dem2002,cmy2009,mlc2022,byd2022,acd2019,rca2022}. Techniques to directly laser cool molecules have developed rapidly in the past decade, with molecular magneto-optical traps (MOTs) demonstrated for several diatomic \cite{nmd2016,twt2017,aad2017,dwy2020} and polyatomic \cite{vhd2022} species. Subsequent sub-Doppler gray molasses cooling to temperatures $\lesssim 50$ $\mu$K \cite{cdt2019,msd2018,dwy2020,cad2018} has enabled loading of molecules into conservative optical dipole traps (ODTs) \cite{cad2018,wbd2021,lhc2022,hvd2022,ljd2021}. Bulk gases of laser cooled molecules in ODTs have been demonstrated with number densities $n_0 \sim10^9 \text{ cm}^{-3}$ and phase space densities $\Phi \sim 10^{-7}$ \cite{cad2018,wbd2021,lhc2022,hvd2022,ljd2021}. However, higher number and phase space densities are needed to implement collisional (evaporative and/or sympathetic) cooling of the trapped molecules, which is likely needed to achieve quantum degeneracy in such systems.

Collisional cooling requires a sufficiently high rate of thermalizing (elastic) collisions \cite{aec1995,sbl2022}. However, experiments with trapped ultracold molecules typically see rapid loss due to inelastic molecule-molecule collisions. Loss mechanisms include chemical reactions, as well as ``sticky collisions" where long-lived collision complexes are formed, which are kicked out of the trap by absorbing a trap light photon or by colliding with a third body \cite{gfc2019,hyw2021,vgo2020,bcl2023,spn2019,cad2020,trn2014,paz2015}. Recent experiments with assembled bi-alkali molecules, at much lower temperatures ($\lesssim 900$ nK), have demonstrated evaporative cooling by suppressing the inelastic collision rate using microwave fields \cite{khu2018,lqu2018,abd2021,sbl2022,lcw2023,bww2023} or static electric fields \cite{gb2016,lty2021}, while enhancing the elastic collision rate. This opens a path towards collisional cooling of molecules, if the density is sufficient to observe collisions. 

For directly laser cooled molecules, inelastic collisions have been reported between pairs of CaF molecules in tweezers \cite{cad2020}, where subsequent microwave shielding was demonstrated \cite{abd2021}, and between molecules and atoms in a magnetic trap \cite{jct2021,spn2019}. Thus far, however, bulk gases of directly laser cooled molecules have been too dilute for either elastic or inelastic molecule-molecule collisions to be observed. There are two primary reasons for this. First, standard red-detuned molecular MOTs (red-MOTs) have low molecule number ($N \lesssim 10^{5}$), due to inefficient slowing of the source molecular beam and low capture velocity of the MOT. Second, transfer efficiency from these red-MOTs into ODTs is low (typically $\lesssim 5$\%) \cite{cad2018,ljd2021}. This is due to sub-Doppler heating from the Type-II transitions ($N_{g}=1\rightarrow N_{e}=0$, where $N_g\{N_e\}$ is the rotational angular momentum of the ground \{excited\} state) required to be driven for rotational closure of molecular optical cycling transitions~\cite{dta2016,dta2018,lde2023}, limiting typical red-MOT radii to $\sigma\gtrsim 1$\,mm and temperatures to $T\gtrsim 1$\,mK~\cite{twt2017,nmd2016,aad2017,cdy2018,vhd2022,lhc2022} after a compression stage. The temperature can be reduced further to $\lesssim 50$ $\mu$K by blue-detuned molasses \cite{cdt2019,msd2018,dwy2020,cad2018}, but this does not provide any spatial compression of the molecular cloud.

 This has led to interest in `blue-detuned' MOTs (blue-MOTs), which can exhibit sub-Doppler cooling while simultaneously maintaining strong confining forces, with Type-II transitions.  This was first demonstrated in Rb atoms~\cite{jdt2018}, and recently shown to work for the specific case of YO (yttrium-monoxide) molecules~\cite{bay2023}. Recently published numerical simulations \cite{lde2023} suggested a more generic method to produce blue-MOTs for a large class of laser-coolable molecules, which should enable efficient transfer of molecules from a MOT to an ODT.

 In this paper, we experimentally realize this novel, generic scheme to produce a blue-MOT of SrF molecules. With it we achieve $\sim10^2$ gain in $n_0$ and $\sim10^4$ gain in $\Phi$ compared to our compressed red-MOT. We load an ODT from this blue-MOT with $\sim30\%$ transfer efficiency, $\sim$ 10x higher than from a compressed red-MOT \cite{ljd2021,cad2018}. With this high density in the ODT, we are able to observe inelastic molecule-molecule collisions that result in two-body loss; to our knowledge this is the first such observation in a bulk gas of directly laser-cooled molecules.

\begin{figure}
    \includegraphics{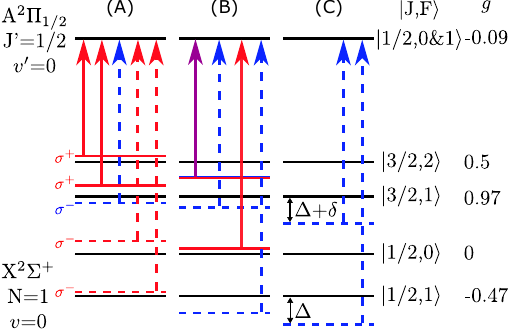}%
    \caption{Relevant SrF level structure and laser-driven transitions for different stages in the experiment, with hyperfine levels ($\ket{J,F}$) and their magnetic g-factors listed. Solid (dashed) lines indicate $\sigma^+ (\sigma^-)$ laser polarization, and color indicates red/blue detuning. (A) Red-MOT, which employs the dual frequency mechanism on $\ket{3/2,1}$. (B) Blue-MOT, where the laser addressing $\ket{3/2,2}$ is now blue, but also provides the red detuning needed for the dual frequency mechanism on $\ket{3/2,1}$ (purple arrow). (C) $\Lambda$-cooling, where only two lasers address $\ket{3/2,1}$ and $\ket{1/2,1}$.}
    \label{fig:level structure}
\end{figure}

Our apparatus is very similar to that used in our prior work \cite{bmd2014,ljd2021}, but with several changes to improve the number of molecules captured in our MOT.  We start with a cryogenic buffer gas beam source (CBGB) \cite{bmd2014}, where SrF molecules are produced by laser ablation of a solid Sr target, with SF$_6$ gas reacting with the ablated Sr to make SrF. The molecules collide with cold (4 K) He gas and exit the cell at forward velocity $\sim 130$ m/s, then are slowed using the white light slowing technique \cite{bsd2012,bmd2014} on the $X\rightarrow B$ transition for 14.5 ms. 

Slowed molecules are captured in a direct current (DC) red-MOT. Here 3 hyperfine levels are addressed by solely red-detuned light, while simultaneous red- and blue-detuned light is applied on the $\ket{J=3/2,F=1}$ state (Fig.~\ref{fig:level structure}(a)) to create a dual-frequency trapping force \cite{tst2015}.  Initially, the per-beam peak laser intensity is $I\sim 100$\,mW/cm$^{2}$ and the axial B-field gradient is $b=16$\,G/cm. After capturing the molecules, we linearly increase the gradient to $b=29$\,G/cm and lower the intensity to $I\sim 10$\,mW/cm$^{2}$ over 30\,ms. In this `compressed' MOT, the cloud radius (in this work, we define radius as the Gaussian r.m.s width unless noted otherwise) is $\sigma\approx1$\,mm, the temperature is $T\approx1$\,mK, and the number of trapped molecules is $N_{\text{red}}\approx 2.5\times 10^{4}$. The molecule number is determined by switching off the gradient and taking a fluorescence image (2 ms exposure) with $I\sim170$ mW/cm$^2$, where the scattering rate is measured using the procedure from \cite{nmd2016} and the detection efficiency is calibrated from off-line measurements \cite{barry}. The fluorescence image is integrated along the radial direction, then fit to a 1D Gaussian plus constant offset; the fluorescence counts are extracted from the Gaussian integral. The temperature is measured using the time-of-flight (TOF) expansion method. We note that in prior work, we began with a radiofrequency (RF) red-MOT \cite{nmd2016,ljd2021}, which traps $\sim30\%$ more molecules with similar size and temperature as the DC red-MOT here. However, switching the B-field from RF to DC for the subsequent blue-MOT configuration is experimentally challenging, and we use the DC red-MOT here instead.

\begin{figure}
    \includegraphics{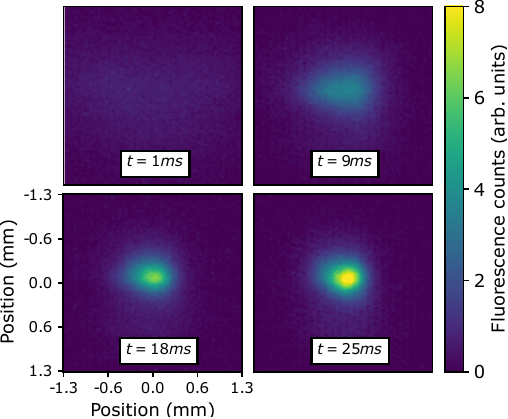}%
    \caption{Fluorescence images showing capture into the blue-MOT as a function of time $t$ after switching from the red-MOT (2 ms in situ exposure starting at time $t$). The loading is quick and efficient, with $\approx$ 80\% of molecules captured within 30 ms of the start of loading.}
    \label{fig:Blue MOT loading}
\end{figure}

Next, we instantaneously jump to the blue-MOT configuration. The laser frequencies are changed to those in Fig.~\ref{fig:level structure}(b), and the intensity is increased to $I\sim 170$\,mW/cm$^{2}$, corresponding to $I/I_{\text{sat}} \sim 60$, where $I_{\text{sat}}$ is the saturation intensity. As in the red-MOT, a dual-frequency scheme is applied to the $|J=3/2,F=1\rangle$ state. However, blue-detuned light is applied to the other $F\neq 0$ states resulting in simultaneous application of both sub-Doppler cooling and spatial confinement \cite{lde2023,bay2023,jdt2018}.

We find that $\sim$80\% of the molecules from the compressed red-MOT are captured by the blue-MOT. Within 30\,ms, the MOT radius along the radial (axial) direction is reduced to as low as $\sigma_{rad}\approx 149 \,\mu$m ($\sigma_{ax}\approx 147\,\mu$m) and the temperatures to as low as $T_{ax}\approx T_{rad} \approx 200\,\mu$K (see Fig.~\ref{fig:Blue MOT loading}), corresponding to a maximum $n_0 \approx 4 \times 10^{8}$\,cm$^{-3}$. The temperature can be lowered further to $T_{ax,rad}\approx60\,\mu$K, by reducing $I$ to 34\,mW/cm$^{2}$, though this results in increasing the radii to $\sigma_{rad}\approx230\,\mu$m ($\sigma_{ax}\approx153\,\mu$m). The blue-MOT reaches a maximum $\Phi \approx 1.6\times 10^{-9}$, a gain of $\sim10^{4}$ compared to the compressed red-MOT. 

We note that our trapping scheme is substantially different from that used for YO molecules in Ref.~\cite{bay2023}, where only blue-detuned light was used.  We were, by contrast, unable to observe trapping without employing a dual-frequency mechanism. We believe the difference lies in the fact that YO, unlike SrF, has a magnetically insensitive ground state hyperfine manifold with $F\neq 0$.  This feature has been observed to increase the robustness of sub-Doppler cooling in magnetic fields~\cite{dwy2020}.  The lack of this feature in SrF (and most other laser cooled molecules) may necessitate the dual-frequency mechanism, which can provide stronger confining forces~\cite{lde2023} at the cost of some heating. Indeed, we observe a stronger restoring force ($\sim$10x faster compression) and smaller minimum trap volume (by a factor of 2) at the cost of higher minimum blue-MOT temperature (60 $\mu$K vs 38 $\mu$K) compared to the pure-blue YO MOT~\cite{bay2023}.

Next, we load the ODT by switching the lasers to the $\Lambda$-enhanced gray molasses \cite{cad2018,ljd2021} configuration in Fig.~\ref{fig:level structure}(c), and turning off the quadrupole field gradient. The ODT details are described elsewhere \cite{ljd2021}; in brief, the ODT is formed from a 53 W, 1064 nm laser focused to a $1/e^2$ radius of 38 $\mu$m, with a trap depth $U_T\approx 1.3$ mK for SrF. We find that loading is optimized for two-photon detuning $\delta = 2\pi\times 0.11$ MHz, one-photon detuning $\Delta = 2\pi\times 22$ MHz, and $I\sim57$ mW/cm$^2$. Owing to the small size of the blue-MOT, the ODT is rapidly loaded, with up to 30\% transfer efficiency achieved within 20 ms. This is an order of magnitude higher efficiency than achieved when loading from type-II red-MOTs \cite{cad2018,ljd2021}. Under optimal conditions, we load $N\approx 4000$ molecules in the ODT, at $T\approx 40$ $\mu$K, and $n_0\approx 3.4\times10^{10} \text{ cm}^{-3}$. We note in passing that here, different from our previous observations, the optimal polarization of the ODT beam is linear and the temperature is higher \cite{ljd2021}. We have so far been unable to trace the source of this change. 

With these starting conditions, we look for evidence of inelastic molecule-molecule collisions. To study collisional loss, we perform measurements of the number of molecules remaining in the trap ($N$) as a function of the hold time ($t_h$). For all of these measurements, we load the ODT for 20 ms, then let untrapped molecules fall out of the trap by turning off the $\Lambda$-cooling light for 32 ms. This defines $t_h=0$. We then measure the remaining number at time $t_h$, either by imaging in-situ with the $\Lambda$-cooling light (for all points $t_h<1$ s) \cite{cad2018}, or by recapturing in the compressed red-MOT and imaging in-situ (for all points $t_h\ge1$ s). The scattering rate for each method is determined by comparing the fluorescence counts to those from a free space image (2 ms exposure) at $I\sim170$ mW/cm$^2$. We assign un-corrrelated uncertainties to each $N(t_h)$ data point by adding in quadrature contributions from fit uncertainties, the shot-to-shot fluctuations in the initial number, and uncertainties in the ratio of the extracted number between the two imaging methods \cite{supp}.

\begin{figure}
    \includegraphics{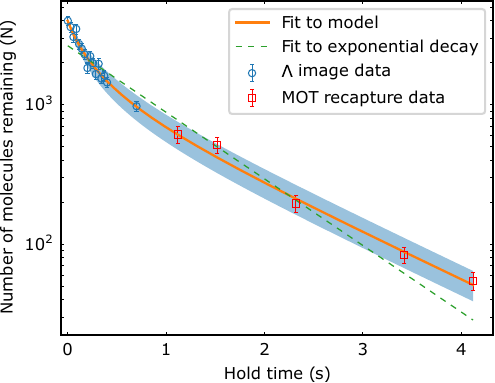}%
    \caption{Number of molecules remaining in the trap as a function of trap hold time. Each point is an average of 15 images, and the error bars account for uncertainties as described in the main text. Data for $t_h<1$ s are $\Lambda$-images (blue circles) and the rest are MOT recapture images (red squares). The data shows a clear deviation from an exponential decay, a classic signature of two-body loss. By fitting to a model where $\sigma_z$ is increasing linearly with time, we extract a two-body loss rate coefficient $\beta=2.7^{+1.2}_{-0.8} \times 10^{-10} \text{ cm}^3 \text{ s}^{-1}$, and a one-body loss rate $\tau=1.3(1)$ s. The shaded area indicates the uncertainty in $\beta$.}
    \label{fig:ODT lifetime curve}
\end{figure}

First, we measure the loss rate in the maximally loaded ODT, with average initial number $N_0\approx4000$. We observe a fast initial loss, followed by a long slow decay, as is characteristic of two-body loss processes (see Fig.~\ref{fig:ODT lifetime curve}). Density dependent losses are modeled using the two-body loss rate equation, with evolution of the number density $n$ given by:
\begin{equation}\label{eq1}
    \dot{n} = -\frac{1}{\tau} n -\beta n^2   ,
\end{equation}

\noindent where $\tau$ is the one-body lifetime and $\beta$ is the two-body loss rate coefficient. To convert eq.~\ref{eq1} to a number evolution, we assume a Gaussian spatial distribution and define an effective volume ($V_{\text{eff}}=(2\sqrt{\pi})^{3} \sigma_x\sigma_y\sigma_z$) occupied by the molecules \cite{trn2014}; here $z$ is along the axial direction of the ODT, and $x$ ($y$) is along the transverse direction in (perpendicular to) the imaging plane. This allows us to integrate over the volume to obtain:
\begin{equation}\label{eq2}
    \dot{N} = -\frac{1}{\tau} N - \frac{\beta}{V_{\text{eff}}} N^2.
\end{equation}

\noindent If the spatial distribution is constant in time, eq.~\ref{eq2} has an analytical solution:
\begin{equation}\label{eq3}
    N(t) = \frac{N_0}{\left(1+\frac{\beta\tau N_0}{V_{\text{eff}}}\right)e^{-t/\tau} - \frac{\beta N_0 \tau}{V_{\text{eff}}}}.
\end{equation}

Our imaging system cannot resolve the transverse radius ($\sigma_{x}$) of the molecular cloud in the ODT. We also cannot observe properties in the $y$-direction. We do directly measure the cloud radius along its axial direction ($\sigma_{z}$), as well as the temperatures $T_{x}$ and $T_{z}$. We then infer $\sigma_{x}$ using the calculated trap depth, measured ODT beam profile, and value of $T_{x}$ \cite{supp}, and assume $\sigma_y = \sigma_x$ by symmetry.

We observe that the measured value of $\sigma_z$ increases from its initial value linearly with hold time. We suspect this results from the ODT beam profile changing due to thermal lensing from optics along the beam path \cite{supp}. We observe an increase in $T_{z}$ consistent with the observed increase in $\sigma_z$. However, we observe no change in $T_{x}$ over time, so we assume that $\sigma_{x}$ (and hence $\sigma_{y}$) does not change. To model this behavior, we treat $V_{\text{eff}}$ as a function of time in eq.~\ref{eq2}, with $\sigma_z$ increasing at the measured rate. We then numerically integrate eq.~\ref{eq2} to find values of $\beta$ and $\tau$ that minimize the reduced chi squared ($\chi_{\text{red}}^2$) of this model. With $N_0=4000$, we find $\beta_{\rm 4K}=2.7(5) \times 10^{-10} \text{ cm}^3 \text{ s}^{-1}$ and $\tau_{\rm 4K}=1.3 (1)$ (with $\chi_{\text{red}}^2=0.99$, see Fig.~\ref{fig:ODT lifetime curve}), where we incorporate the uncertainty in $V_{\text{eff}}$ by adding it in quadrature to the uncertainty determined from the fit.

The final extracted value of $\beta$ is strongly dependent on the initial number, so we also consider systematic uncertainties in determining $N_0$. The scattering rate is affected by uncertainty in the vibrational branching ratio $\ket{A^2\Pi_{1/2},v=0}\rightarrow \ket{X^2\Sigma,v=3}$ for SrF \cite{nmd2016,NLEDM2019,supp}, and in the calibration of the imaging optics. We estimate a combined uncertainty of $25\%$ in $N_0$ \cite{supp}. We emphasize that this is different from shot-to-shot fluctuations, and instead is a correlated uncertainty for all points, which in turn leads to an uncertainty in the overall normalization of $\beta$. To determine the effect of this scale uncertainty, we use the same analysis method with initial numbers $N_0=\{3000,5000\}$ (corresponding to the lower and upper bounds given the uncertainty), and numerically integrate eq.~\ref{eq2} to find the optimal $\beta$ for each $N_0$. The final uncertainty for $\beta$ is then assigned as the quadrature sum of contributions from this systematic uncertainty and from the fit error for $N_0=4000$. Finally, we find $\beta = 2.7^{+1.2}_{-0.8} \times 10^{-10} \text{ cm}^3 \text{ s}^{-1}$ and $\tau=1.3(1)$ s.

As a cross-check, we also fit the data to the analytical solution (eq.~\ref{eq3}) by following the prescription from Ref.~\cite{lcw2023}. That is: we first extract $\tau$ by fitting a pure exponential decay to only late-time ($t_h \ge 1$ s) data points, and find $\tau_{\rm l}=1.2(2)$ s. Then, we extract $\beta$ by fixing this value of $\tau$ and fitting only to early-time data points ($t_h < 250$ ms) where the axial radius change is less than 15\%, such that $V_{\text{eff}}$ can be treated as a constant; we use the average $V_{\text{eff}}$ for $t_h<250$ ms. Throughout, we perform the same error analysis as before, and find $\beta_{\rm e} = 2.7^{+1.4}_{-1.0} \times 10^{-10} \text{ cm}^3 \text{ s}^{-1}$ (with $\chi_{\text{red}}^2=1.20$), consistent with results from the more complete model.

\begin{figure}
    \includegraphics{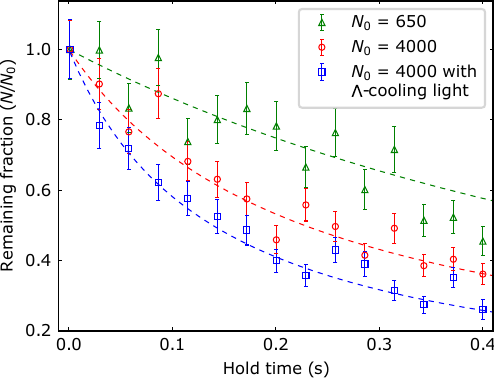}%
    \caption{Short-time evolution of trap population for different starting conditions. Dashed lines are fits for the first 9 points to the two body loss rate model with fixed $\tau=1.3$ s and the average $V_{\text{eff}}$ for $t_h<250$ ms. Data with initial ODT number $N_0\approx650$ (green triangles) has a slower initial loss as compared to the case with $N_0\approx 4000$ (red circles), clearly demonstrating the density dependent loss. The presence of $\Lambda$-cooling light leads to additional two-body loss (blue squares) due to light-assisted collisions, while the one-body loss rate remains the same (as seen in longer-time data, not shown).}
    \label{fig:ODT lifetime curve comparison}
\end{figure}

To further verify the presence of a density-dependent loss, we load the ODT with lower initial number, $N_0 \approx 650$, while keeping the temperature and trap depth the same, thereby reducing the starting density by a factor of 6. This is done by reducing the length of the slowing pulse after laser ablation from 14.5 ms to 9.3 ms. We then perform the same sequence of measurements, and see that the short-time loss rate is reduced (see Fig.~\ref{fig:ODT lifetime curve comparison}), as expected since the initial collision-induced loss time scale $\tau_c\propto 1/\beta n_0$.

There are numerous possible loss channels in our experiment. The molecules are all in the rotational $N=1$ states, and rotational quenching to $N=0$ can lead to large inelastic losses \cite{jct2021}. They also occupy all of the many individual sublevels in the $N=1$ manifold of hyperfine and spin-rotation states; this opens up \textit{p}-wave and \textit{f}-wave collision channels that would be absent if all the (bosonic) molecules were in the same quantum state. In addition, colliding pairs of SrF molecules can undergo the barrierless chemical reaction SrF + SrF $\rightarrow$ SrF$_2$ + Sr \cite{mb2011}. Finally, sticky collisions between the molecules can also lead to losses \cite{gfc2019,hyw2021,vgo2020,bcl2023,spn2019,cad2020,trn2014,paz2015}. 

We compare our measured value of $\beta$ to some experimental and theoretical benchmarks. The molecules in the trap are at temperatures above the \textit{p}- and \textit{d}-wave barriers ($\approx 5$ $\mu$K and $\approx 30$ $\mu$K respectively) determined by the van der Waals $C_6$ coefficient for $N=1$ states in SrF \cite{supp}. We calculate the thermally averaged unitarity limit for $\beta$ by assuming the molecules in the trap obey a Maxwell-Boltzmann distribution of velocities and that the probability for loss is 0 (1) for kinetic energies below (above) the barrier. We find the unitarity limit $\beta_{\text{max}} \approx 11 \times 10^{-10} \text{ cm}^{3} \text{ s}^{-1}$ \cite{ld2017,supp}, $\approx 4$x above the experimental value. Values of $\beta$ of the same order as that measured by us have been reported in Refs.~\cite{gfc2019,hyw2021,vgo2020,bcl2023,trn2014,paz2015,cad2020,spn2019,jct2021}. In most of these experiments, the molecules are in a single quantum state, or are at much colder temperatures, making a direct comparison infeasible. The closest case to our conditions was in the observation of collisions between a pair of CaF molecules in a mixture of $N=1$ states in an optical tweezer trap \cite{acd2019}. There, the molecules were at $T\approx 80 \,\mu$K, above the \textit{p}-wave, but below the \textit{d}-wave barrier for CaF ($\approx 20\,\mu$K and $\approx 105\,\mu$K respectively). The reported loss rate coefficient is $\beta_{\text{CaF}} = 40\times10^{-10} \text{ cm}^3 \text{ s}^{-1}$, $\approx 3$x larger than the corresponding unitarity limit $\beta_{\text{max,CaF}}\approx 13\times 10^{-10} \text{ cm}^3 \text{ s}^{-1}$.

We also explore light-assisted collisions due to the $\Lambda$-cooling light. For this, we turn on the $\Lambda$-cooling light at $t_h=0$. We observe that the one-body lifetime is unaffected by the light, however, the two-body loss rate coefficient is increased due to light-assisted collisions. Under these conditions, we find $\beta_{\text{tot}}=4.9^{+1.7}_{-1.2} \times 10^{-10} \text{ cm}^3 \text{ s}^{-1}$. This is two orders of magnitude lower than previously reported using a pair of CaF molecules in an optical tweezer \cite{acd2019}. The combined loss rate coefficient sets an upper bound on the peak density achievable by loading an ODT using $\Lambda$-cooling. Given the typical loading time (20 ms) from the blue-MOT, this bound is $n_0^{\text{max}}\sim 10^{11} \text{ cm}^{-3}$. While the peak densities we achieve are lower than $n_0^{\text{max}}$, it may be possible to reach it if larger numbers of molecules \cite{lde2023}, lower temperatures \cite{ljd2021,wbd2021}, and/or deeper traps can be achieved. We also note that the observed value of $\beta_{\text{tot}}$ is low enough such that light-induced losses during the in-situ imaging with $\Lambda$-cooling light do not substantially affect the extracted values of $N(t_h)$.

In conclusion, we have demonstrated high efficiency loading of a molecular gas into an ODT from a blue-MOT and observed inelastic collisions in a bulk gas of directly laser cooled molecules for the first time. Our results suggest the possibility of using a shielding mechanism to enhance the elastic collision rate while suppressing two-body losses, as already used for evaporative cooling in experiments using assembled bi-alkali molecules \cite{khu2018,lqu2018,abd2021,sbl2022,lcw2023,bww2023,gb2016,lty2021}. Current efforts are underway to prepare the molecules in a single quantum state and then to implement
microwave shielding in our system. This will open a clear path to collisional cooling via evaporation or sympathetic cooling with co-trapped atoms.

We gratefully acknowledge support from AFOSR MURI and the University of Chicago.

\bibliography{bibtexMasterFileDeMille}
\clearpage
\onecolumngrid
\section{Supplemental Material for High density loading and collisional loss of laser cooled molecules in an optical trap}
\twocolumngrid
\setcounter{equation}{0}
\setcounter{figure}{0}
\setcounter{table}{0}
\setcounter{page}{1}
\makeatletter
\renewcommand{\theequation}{S\arabic{equation}}
\renewcommand{\thefigure}{S\arabic{figure}}

\section{Simulations of blue-detuned MOTs}

We have developed software to simulate magneto-optical trapping of molecules in which acceleration $a(z,v_{z})$ is determined as a function of position $z$ and velocity $v_{z}$; although we consider displacement and velocity along only one axis ($z$), the simulation is three-dimensional, and includes the effect of all 6 laser passes and the 3D anti-Helmholtz quadrupole magnetic field.  This has been used previously to simulate `two-color' blue-MOTs where both $X\Sigma\rightarrow A\Pi$ and $X\Sigma\rightarrow B\Sigma$ electronic transitions are driven~\cite{lde2023}.  We used this same software in order to determine the feasibility of a `one-color' MOT, using just the $X\Sigma\rightarrow A\Pi$ transition, but with a dual-frequency trapping mechanism, before we attempted it in the experiment.  

\begin{figure}
    \includegraphics[width=\linewidth]{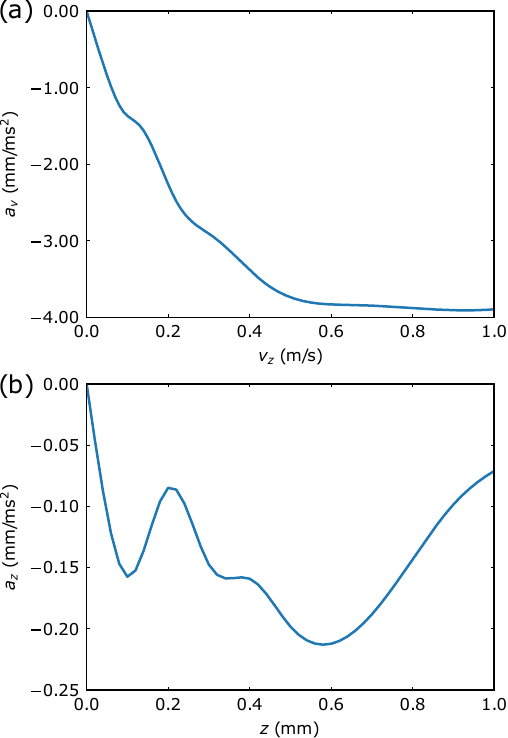}%
    \caption{Simulation results from the OBE solver for the blue-MOT. (a) Acceleration ($a_v$) vs velocity ($v_z$) showing a damping force and (b) acceleration ($a_z$) vs position ($z$) showing a spatial restoring force.}
    \label{blue mot simulation}
\end{figure}

In Fig.~\ref{blue mot simulation}, we show the results of the simulation for the parameters used in the experiment.  Fig.~\ref{blue mot simulation}(a) shows $a(v_{z})=\int_{-zmin}^{zmin} a(z,v_{z})dz$ and Fig.~\ref{blue mot simulation}(b) shows $a(z)=\int_{-vmin}^{vmin}a(z,v_{z})dv_{z}$, where $z_{min}=0.4$\,mm and $v_{min}=0.25$\,m/s are chosen because they represent 2$\sigma$ and 2$v_{T}$, where $\sigma\sim 200\mu$m is the MOT radius observed in the experiment and $v_{T}=\sqrt{k_{B}T/m}$ is the thermal velocity for $T=200\mu$K.  We observe both a sub-Doppler velocity damping force as well as a spatial restoring force, consistent with the observation of trapping in the experiment.  Furthermore, the simulation indicates a spring constant of $\kappa\sim 7\times 10^{-20}$\,N/m (here we approximate the spring constant by taking $\kappa\sim\frac{ma(\textrm{z=400}\mu\textrm{m})}{\textrm{400}\mu\textrm{m}}$), in line with what we observe in the experiment (using the equipartition theorem, $\kappa=\frac{k_{B}T}{\sigma^{2}}$).

The maximum spatial restoring force is roughly an order of magnitude lower than in the previously simulated two-color MOTs~\cite{lde2023}, and so it is possible that two-color trapping can lead to even more gains in MOT density and also enhance the robustness of the blue-MOT to day-to-day laser-pointing fluctuations, which we observed that the blue-MOT was sensitive to.

\section{Lifetime data taking procedure and error analysis}
To get an accurate measurement of the two-body loss rate, it is imperative to get early time data when the molecule number is high and the loss is mostly density dependent, especially in our system where the vacuum limited lifetime is short ($\tau\approx 1.3$ s). As described in the main text, after loading the ODT for 20 ms, we let untrapped molecules fall out of the trap. The earliest time where we can distinguish the trapped molecules from the untrapped falling molecules is $t_{\text{fall}}=32$ ms, where the untrapped molecules are not yet out of the field of view of the camera, but have fallen enough that the trapped molecules can be distinguished by an appropriate choice of the region of interest (ROI). We define $t_h=0$, the starting point of the lifetime measurement to be after this fall period. 

We also divide our data in two chunks. For $t_h< 1$ s, where the number is high and the signal to noise ratio is good, we take a $\Lambda$-image of the molecular fluorescence by exposing the camera for 10 ms while $\Lambda$-cooling light (at $I\sim57$ mW/cm$^2$) is applied to molecules in the ODT. For the rest, where the number is low, we turn off the ODT and recapture the remaining molecules into the compressed red-MOT and take an image of the molecular fluorescence by exposing the camera for 50 ms (at $I\sim10$ mW/cm$^2$). This method can only be used when $t_{\text{fall}} \ge$ 152 ms, in order to not recapture molecules untrapped by the ODT. For each method, we determine the scattering rate by comparing the fluorescence counts of an image taken immediately after $t_{\text{fall}}=32 ( 152)$ ms  (with maximum molecule number) to the counts from a 2 ms exposure image at laser intensity $I\sim170$ mW/cm$^2$, where the scattering rate is measured.

The $\Lambda$-image data is taken as follows. We first generate a list of hold times, and randomize this list. For each hold time $t_i$ in this list, we take a set of 30 images, 15 each for $t_h=t_i$ and for $t_h=0$. The order of these 30 images is randomized as well. We then extract the fluorescence counts as described in the main text. By comparing the values of $N_0$ inferred from the $t_h = 0$ data in each set of data at different values of $t_i$, we are able to measure the shot-to-shot drifts in starting number over the duration of the entire data set. We find that the various $N_0$ values have a standard deviation $\sim 8\%$, and we add this in quadrature to the other uncertainties in the number at each data point. 

For the MOT recapture data, we follow the same procedure, with the additional drop time added. In addition, we take a set of images at a few intermediate $t_h$ using both methods to compare the extracted number and we find that the ratio of the number extracted from MOT recapture to the number from $\Lambda$-images is $N_{MOT recap}/N_{\Lambda-image} = 1.01 \pm 0.10$. We hence include an additional 10\% uncertainty in molecule number for all points with $t_h\ge 1$ s, where the MOT recapture method is used.

In addition to these uncertainties, there is also ambiguity in the determination of the overall scattering rate, because of differing reported values of the branching ratio $\ket{A^2\Pi_{1/2},J=1/2^+,v=0}\rightarrow \ket{X^2\Sigma,N=1,v=3}$ for SrF molecules \cite{nmd2016,NLEDM2019}. This results in an overall scale factor uncertainty in the molecule number when converting from fluorescence counts to number. To account for this, we use the average of the branching ratios from Refs.~\cite{nmd2016,NLEDM2019}, and half their difference as its uncertainty. This leads to $\sim 17\%$ uncertainty in the determination of the starting number. We further take into account uncertainties in the calibration of the light collection optics for the imaging setup ($\sim 10\%$) and we find a combined uncertainty $\sim25\%$ in the overall starting number. We emphasize that this is different from shot-to-shot fluctuations, as this uncertainty affects each data point in the same direction.

\section{$V_{\text{EFF}}$ CALCULATION}
To calculate $V_{\text{eff}}$, we assume that the trap is harmonic and the spatial density is given by:
\begin{equation}
    n(\textbf{r}) = n_0 \; \text{exp}\left(-\frac{x^2}{2\sigma_x^2} \right)\; \text{exp}\left(-\frac{y^2}{2\sigma_y^2} \right)\; \text{exp}\left(-\frac{z^2}{2\sigma_z^2} \right).
\end{equation}
The effective volume is given by $V_{\text{eff}} = \int n(\textbf{r})d^3\textbf{r} / n_0=(2\sqrt{\pi})^3\sigma_x\sigma_y\sigma_z$. We do not have enough resolution to measure the transverse radius $\sigma_x$, and our observations give no direct information about $\sigma_y$. However, we have measured the ODT laser beam profile, and find a good fit to a Gaussian with $1/e^2$ intensity radius $\omega_0=38 (3)$ $\mu$m. We have also calculated the ODT trap depth to be $U_T=1.3 (1)$ mK \cite{ljd2021}, and we measure the transverse temperature of molecules in the ODT using time-of-flight (TOF) expansion technique to be $T_x=40(3)$ $\mu$K. From this we deduce the transverse radius $\sigma_x= \sqrt{\omega_0^2T/4U_T}=3.3(4)$ $\mu$m, and assume $\sigma_y = \sigma_x$ 

\begin{figure}
    \includegraphics[width=\linewidth]{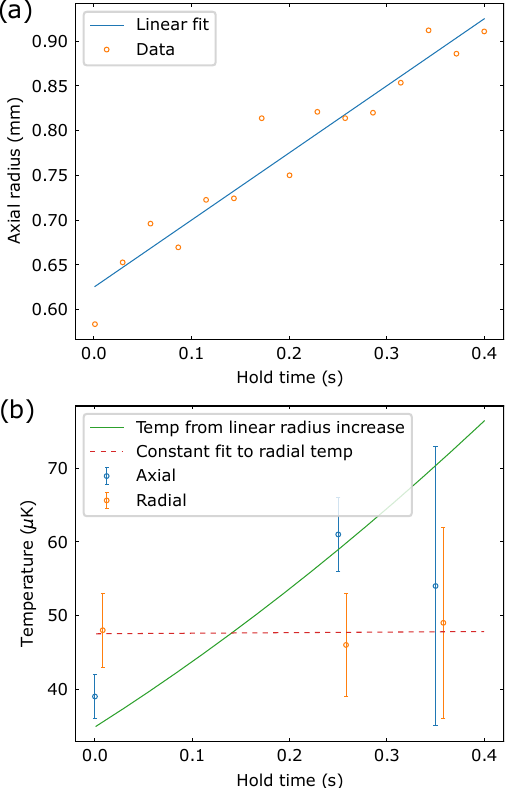}%
    \caption{(a) Axial radius as a function of hold time, and best fit line with $\sigma_{z} = mt_h + c$. We find $m=0.75$ mm/s and $c=0.63$ mm. (b) Axial and radial temperature versus hold time. Radial temperatures are slightly offset in time for clarity. The axial temperature increases as expected from the axial radius increase, while the radial temperature does not noticeably increase.}
    \label{axial width vs time}
\end{figure}

We are able to directly measure the axial radius of the molecular cloud in the ODT, $\sigma_z$, and (as stated in the main text) we observe that $\sigma_z$ increases with time. We model this as a linear increase (see Fig.~\ref{axial width vs time}). We then use the measured value of $\sigma_z$ to deduce the axial temperature of the molecules in the ODT, $T_z^\sigma$, by following the procedure outlined in \cite{ljd2021}, where $T_z^\sigma=2U_T\sigma_z^2/z_R^2$, where $z_R$ is the Rayleigh length of the trap. We find that this inferred axial temperature increase is consistent with the directly measured increase in $T_z$ (Fig.~\ref{axial width vs time}(a)). This justifies our assumption of a linearly increasing value of $\sigma_z$ away from its starting value. We do not see any increase in the measured radial temperature; thus we model $\sigma_{x,y}$ to be constant. We then numerically integrate eq.~\ref{eq2}, while allowing the effective volume to increase as $V_{\text{eff}}(t)=(2\sqrt{\pi})^3\sigma_x\sigma_y\sigma_z(t)$, to find the result described in the main text. To fit to the analytical solution (eq.~\ref{eq3}), we use the average axial radius for $t_h<250$ ms, $V_{\text{eff}} = 3.4 (9) \times 10^{-7} \text{ cm}^{3}$.

\section{Calculation of Van der Waals $C_6$ Coefficient for $\ket{N = 1}$ States}
To determine the centrifugal barriers for two-body SrF scattering, we need to compute the van der Waals (vdW) $C_6$ coefficient that arises from second-order dipolar coupling. Throughout, we assume the molecules are in the ground vibronic manifold $\ket{X^2\Sigma^+, v = 0}$, and for simplicity we ignore electronic and nuclear spins.

For rotational ground state $\ket{N = 0}$ molecules, the rotational wavefunction is spatially isotropic and only a single rotational sublevel is occupied. This leads to the well-known result \cite{qbpk2011} $C_6^g = -\left[ 1/(4\pi\epsilon_0)^2\right]\, \cdot \,d_0^4/(6B_0)$, where $d_0 = 3.47$ Debye is the ground state permanent dipole moment of SrF and $B_0 = 2\pi\hbar\times 7.5$ GHz is the ground state rotational constant of SrF. However, our molecules are in an incoherent mixture of $\ket{N = 1}$ states, which (in the absence of external fields, and ignoring effects due to spin-rotation and hyperfine couplings) comprise a nine-fold degenerate subspace in the space of $\ket{N = 1}$ two-body states. Following the approach of \cite{ld2017}, we apply second-order degenerate perturbation theory on the intermolecular potential operator $\hat{V}_{\text{AB}}$ in order to obtain the $\ket{N = 1}$ $C_6$ coefficients.

\begin{figure}
    \includegraphics[width=\linewidth]{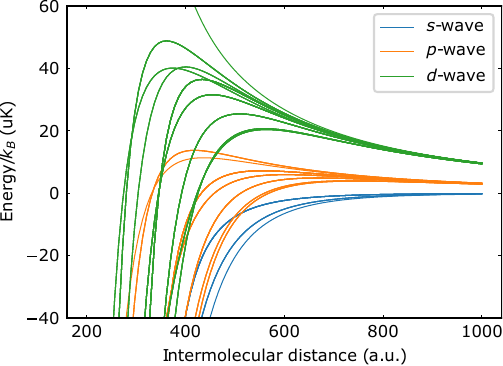}%
    \caption{Numerically computed potential energy curves in a single channel model for two-body SrF scattering, where colliding molecules are in the $\ket{N = 1}$ manifold. Curves up to the $\ell = 2$ partial wave are shown.}
    \label{vdW curves}
\end{figure}

In a body-fixed (BF) frame where the orientation of the vector between the two molecules is fixed, the resultant $C_6$ coefficients, as a function of $d_0$ and $B_0$, are listed in Table 6 of \cite{ld2017}. We used these values in eq.\@ \ref{eq:barTemp} to compute the centrifugal barrier heights. Due to the anisotropic nature of the vdW interaction, the $C_6$ coefficients including the relative angular motion of the molecules in the space-fixed (SF) frame must be accounted for. To find $C_6$ values in the SF frame, we numerically compute matrix elements of the second-order degenerate perturbation operator $\hat{W}_{\text{AB}}^{\text{SF}}$ associated with $\hat{V}_{\text{AB}}$, as given by eq.\@ 82 of \cite{ld2017}. By incorporating the $\ell^\text{th}$ partial wave $\ket{\ell, m_\ell}$ into that matrix element, and subsequently diagonalizing the combined potential $\hat{W}_{\text{AB}}^{\text{SF}} + \hat{\ell}^2/(2\mu R^2)$ (where $R$ is the intermolecular separation) in the subspace of $\ket{N = 1}$ two-body states, we obtain the intermolecular potential curves shown in Fig. \ref{vdW curves}. The resultant barrier heights that we obtain from these curves are in nearly exact agreement with those obtained from using the BF calculation results. Hence, for computational ease we use the analytically determined BF centrifugal barriers in the rest of this work.

\section{Unitarity Limit Calculation}
Here, we compute the unitarity limit for two-body scattering of SrF molecules in an incoherent mixture of $\ket{N = 1}$ states. The unitarity limit corresponds to the maximum possible loss allowed by quantum scattering theory. Since our molecules are not in a single quantum state, all partial waves (indexed by $\ell$) contribute to the unitarity limit scattering cross section $\sigma$. In this limit, the scattering cross section of the $\ell^\text{th}$ partial wave is:
\begin{equation}
    \sigma_\ell = \frac{4\pi (2\ell + 1)}{k^2}.
\end{equation}
As usual in two-body scattering, we work in the center-of-mass frame of the two-body system. Therefore, the wavevector $\vec{k}$ relates to the collisional energy $E_r$ and relative velocity $\vec{v}_r$ of the two particles via $E_r = \hbar^2 k^2/2\mu$ and $\vec{v}_r = \hbar \vec{k}/\mu$, respectively, where $\mu$ is the reduced mass. For two-body SrF scattering, $\mu = M/2$ where $M \approx 107$ amu is the mass of SrF.

We assume that SrF molecules in our trap obey a Maxwell-Boltzmann (MB) thermal distribution in their absolute velocities and energies. Since the convolution of two Gaussians is another Gaussian, it follows that the probability density function $f_r(v_r, T)$ of relative speeds for two-body SrF scattering is also a normalized MB distribution:
\begin{equation}
    f_r(v_r, T)\,dv_r = \sqrt{\frac{2}{\pi}}\left(\frac{\mu}{k_B T}\right)^{\frac{3}{2}}v_r^2\exp{\left(-\frac{\mu v_r^2}{2k_B T}\right)}\, dv_r,
\end{equation}
where $\int_0^\infty f_r(v_r, T)\, dv_r = 1$ and $T$ is the initial temperature of molecules in our trap. 

We define a thermally averaged two-body loss rate coefficient $\beta_\text{th}(\ell, T)$ for the $\ell^\text{th}$ partial wave by considering which collisional speeds will allow for unitary loss. We make the simplifying assumption that unitary loss occurs with unit (zero) probability when $E_r$ is greater (less) than the centrifugal barrier of the intermolecular potential, i.e. we assume that colliding SrF molecules cannot tunnel through the centrifugal barrier. By denoting the collisional speed associated to the barrier height as $v_b$, we have:
\begin{equation}
    \beta_\text{th}(\ell, T) = \int_{v_b}^\infty f_r(v_r, T)\sigma_\ell(v_r) v_r\,dv_r.
\end{equation}
By introducing the dimensionless parameter $x \equiv E_r/k_B T$ and substituting in the expressions for $f_r(v_r, T)$ and $\sigma_\ell(v_r)$, the expression for $\beta_\text{th}(\ell, T)$ can be simplified as:
\begin{equation}
    \beta_\text{th}(\ell, T) = \frac{2\pi\hbar^2(2\ell + 1)}{\mu k_B T}\sqrt{\frac{8 k_B T}{\pi \mu}}\int_{T_b(\ell)/T}^\infty e^{-x}\,dx.
\end{equation}
We identify $\lambda_\text{th} = \sqrt{2\pi\hbar^2/\mu k_B T}$ and $\bar{v}_\text{th} = \sqrt{8k_B T/\pi \mu}$ as the thermal de Broglie wavelength and average speed, respectively, of a particle with mass $\mu$ in an ensemble at temperature $T$ that obeys MB statistics. Therefore, we conclude:
\begin{equation}
    \beta_\text{th}(\ell, T) = \lambda_\text{th}^2 \bar{v}_\text{th}(2\ell + 1)e^{-T_b(\ell)/T},
\end{equation}
where $k_B T_b(\ell)$ is the height of the centrifugal barrier experienced by the $\ell^\text{th}$ partial wave.

Each distinct vdW $C_6$ coefficient corresponds to a distinct value for $T_b(\ell)$. For the $i^\text{th}$ $C_6$ coefficient, denoted $C_{6,i}$, we relate it to the $i^\text{th}$ barrier $T_{b,i}(\ell)$ as follows. If we neglect quadrupole-quadrupole interactions, the leading order terms in the SrF $\ket{N = 1}$ + SrF $\ket{N = 1}$ intermolecular potential lead to the potential:
\begin{equation}
    V_i(r) = \frac{\hbar^2 \ell(\ell+1)}{2\mu r^2} + \frac{C_{6,i}}{r^6}.
\end{equation}
If $C_{6,i} > 0$, the vdW interaction is repulsive and no barrier exists. We treat this as meaning the molecules never reach short range, and so the contribution to $\beta_\text{th}(\ell, T)$ here is zero. If $C_{6,i} < 0$, the vdW interaction is attractive and there will exist a maximum in $V_i(r)$ at $r_\text{b}$, corresponding to the centrifugal barrier. By only considering cases where $C_6 < 0$, we analytically compute the barrier height of the $\ell^\text{th}$ partial wave to be:

\begin{equation}
    T_{b,i}(\ell) = \frac{V(r_\text{b})}{k_B} = \left(\frac{\hbar^2 \ell(\ell+1)}{\mu}\right)^{\frac{3}{2}}\left(\frac{1}{54 |C_{6,i}|}\right)^{\frac{1}{2}}\frac{1}{k_B}.\label{eq:barTemp}
\end{equation}

For each two-body eigenstate, we compute their barrier heights up to the $h$-wave contribution ($\ell = 5$). We neglect summation over all partial waves with $\ell > 5$ because their contribution to $\beta_\text{th}(\ell, T)$ is increasingly exponentially suppressed. We thus obtain the total $\beta_\text{th,i}(T)$ for the $i^\text{th}$ two-body eigenstate:
\begin{equation}
    \beta_\text{th,i}(T) = 
    \begin{cases}
        \sum_{\ell = 0}^5 \lambda_\text{th}^2 \bar{v}_\text{th}(2\ell + 1)e^{-T_{b,i}(\ell)/T} & \text{if } C_6 < 0\\[3pt]
        0 & \text{if } C_6 > 0.
    \end{cases}
\end{equation}

We finally obtain the overall two-body loss rate coefficient in the unitarity limit for an incoherent mixture of $\ket{N = 1}$ states, denoted as $\langle \beta_\text{th}(T)\rangle_{\ket{N = 1}}$, by taking a statistical average over all nine possible $\ket{N = 1}$ two-body eigenstates. Here, we assume a uniform probability distribution over all possible states, i.e.\@ $p_i = 1/9$ $\forall i \in \{1, 2, ..., 8, 9\}$. So, we have:
\begin{equation}
    \langle \beta_\text{th}(T)\rangle_{\ket{N = 1}} = \sum_{i = 1}^9 p_i\, \beta_\text{th,i}(T).
\end{equation}
Carrying out this computation, we find $\langle \beta_\text{th}(T)\rangle_{\ket{N = 1}} \approx 11 \times 10^{-10}$ cm$^3$/s at $T = 40 \,\mu$K, the temperature of SrF molecules in our ODT.

We note that the intermolecular potential for $\ket{N = 1}$ states contains a first-order contribution from the quadrupole-quadrupole interaction, of the form $C_5/r^5$ \cite{ld2017}. Here, $C_5 \sim \theta_{zz}^2/(4\pi\varepsilon_0)$, where $\theta_{zz} = 8.95\, ea_0^2$ is the quadrupole moment of SrF \cite{mb2011}. This extra term affects the barrier heights only minimally; by including the quadrupole-quadrupole interaction, we found that $\langle \beta_\text{th}(T)\rangle_{\ket{N = 1}}$ is changed by only $\approx 3$\%, and hence its effect is negligible..

\bibliography{bibtexMasterFileDeMille}

\begin{thebibliography}{51}%
\makeatletter
\providecommand \@ifxundefined [1]{%
 \@ifx{#1\undefined}
}%
\providecommand \@ifnum [1]{%
 \ifnum #1\expandafter \@firstoftwo
 \else \expandafter \@secondoftwo
 \fi
}%
\providecommand \@ifx [1]{%
 \ifx #1\expandafter \@firstoftwo
 \else \expandafter \@secondoftwo
 \fi
}%
\providecommand \natexlab [1]{#1}%
\providecommand \enquote  [1]{``#1''}%
\providecommand \bibnamefont  [1]{#1}%
\providecommand \bibfnamefont [1]{#1}%
\providecommand \citenamefont [1]{#1}%
\providecommand \href@noop [0]{\@secondoftwo}%
\providecommand \href [0]{\begingroup \@sanitize@url \@href}%
\providecommand \@href[1]{\@@startlink{#1}\@@href}%
\providecommand \@@href[1]{\endgroup#1\@@endlink}%
\providecommand \@sanitize@url [0]{\catcode `\\12\catcode `\$12\catcode
  `\&12\catcode `\#12\catcode `\^12\catcode `\_12\catcode `\%12\relax}%
\providecommand \@@startlink[1]{}%
\providecommand \@@endlink[0]{}%
\providecommand \url  [0]{\begingroup\@sanitize@url \@url }%
\providecommand \@url [1]{\endgroup\@href {#1}{\urlprefix }}%
\providecommand \urlprefix  [0]{URL }%
\providecommand \Eprint [0]{\href }%
\providecommand \doibase [0]{https://doi.org/}%
\providecommand \selectlanguage [0]{\@gobble}%
\providecommand \bibinfo  [0]{\@secondoftwo}%
\providecommand \bibfield  [0]{\@secondoftwo}%
\providecommand \translation [1]{[#1]}%
\providecommand \BibitemOpen [0]{}%
\providecommand \bibitemStop [0]{}%
\providecommand \bibitemNoStop [0]{.\EOS\space}%
\providecommand \EOS [0]{\spacefactor3000\relax}%
\providecommand \BibitemShut  [1]{\csname bibitem#1\endcsname}%
\let\auto@bib@innerbib\@empty
\bibitem [{\citenamefont {DeMille}(2002)}]{dem2002}%
  \BibitemOpen
  \bibfield  {author} {\bibinfo {author} {\bibfnamefont {D.}~\bibnamefont
  {DeMille}},\ }\bibfield  {title} {\bibinfo {title} {Quantum computation with
  trapped polar molecules},\ }\href@noop {} {\bibfield  {journal} {\bibinfo
  {journal} {Phys. Rev. Lett.}\ }\textbf {\bibinfo {volume} {88}},\ \bibinfo
  {pages} {067901} (\bibinfo {year} {2002})}\BibitemShut {NoStop}%
\bibitem [{\citenamefont {Carr}\ \emph {et~al.}(2009)\citenamefont {Carr},
  \citenamefont {DeMille}, \citenamefont {Krems},\ and\ \citenamefont
  {Ye}}]{cmy2009}%
  \BibitemOpen
  \bibfield  {author} {\bibinfo {author} {\bibfnamefont {L.~D.}\ \bibnamefont
  {Carr}}, \bibinfo {author} {\bibfnamefont {D.}~\bibnamefont {DeMille}},
  \bibinfo {author} {\bibfnamefont {R.~V.}\ \bibnamefont {Krems}},\ and\
  \bibinfo {author} {\bibfnamefont {J.}~\bibnamefont {Ye}},\ }\bibfield
  {title} {\bibinfo {title} {Cold and ultracold molecules: science, technology
  and applications},\ }\href@noop {} {\bibfield  {journal} {\bibinfo  {journal}
  {New J. Phys.}\ }\textbf {\bibinfo {volume} {11}},\ \bibinfo {pages} {055049}
  (\bibinfo {year} {2009})}\BibitemShut {NoStop}%
\bibitem [{\citenamefont {Holland}\ \emph {et~al.}(2022)\citenamefont
  {Holland}, \citenamefont {Lu},\ and\ \citenamefont {Cheuk}}]{mlc2022}%
  \BibitemOpen
  \bibfield  {author} {\bibinfo {author} {\bibfnamefont {C.~M.}\ \bibnamefont
  {Holland}}, \bibinfo {author} {\bibfnamefont {Y.}~\bibnamefont {Lu}},\ and\
  \bibinfo {author} {\bibfnamefont {L.~W.}\ \bibnamefont {Cheuk}},\ }\bibfield
  {title} {\bibinfo {title} {On-demand entanglement of molecules in a
  reconfigurable optical tweezer array},\ }\href@noop {} {\bibfield  {journal}
  {\bibinfo  {journal} {arXiv:2210.06309v1}\ } (\bibinfo {year}
  {2022})}\BibitemShut {NoStop}%
\bibitem [{\citenamefont {Bao}\ \emph {et~al.}(2022)\citenamefont {Bao},
  \citenamefont {Yu}, \citenamefont {Anderegg}, \citenamefont {Chae},
  \citenamefont {Ketterle}, \citenamefont {Ni},\ and\ \citenamefont
  {Doyle}}]{byd2022}%
  \BibitemOpen
  \bibfield  {author} {\bibinfo {author} {\bibfnamefont {Y.}~\bibnamefont
  {Bao}}, \bibinfo {author} {\bibfnamefont {S.~S.}\ \bibnamefont {Yu}},
  \bibinfo {author} {\bibfnamefont {L.}~\bibnamefont {Anderegg}}, \bibinfo
  {author} {\bibfnamefont {E.}~\bibnamefont {Chae}}, \bibinfo {author}
  {\bibfnamefont {W.}~\bibnamefont {Ketterle}}, \bibinfo {author}
  {\bibfnamefont {K.~K.}\ \bibnamefont {Ni}},\ and\ \bibinfo {author}
  {\bibfnamefont {J.~M.}\ \bibnamefont {Doyle}},\ }\bibfield  {title} {\bibinfo
  {title} {Dipolar spin-exchange and entanglement between molecules in an
  optical tweezer array},\ }\href@noop {} {\bibfield  {journal} {\bibinfo
  {journal} {arXiv:2211.09780v1}\ } (\bibinfo {year} {2022})}\BibitemShut
  {NoStop}%
\bibitem [{\citenamefont {Anderegg}\ \emph {et~al.}(2019)\citenamefont
  {Anderegg}, \citenamefont {Cheuk}, \citenamefont {Bao}, \citenamefont
  {Burchesky}, \citenamefont {Ketterle}, \citenamefont {Ni},\ and\
  \citenamefont {Doyle}}]{acd2019}%
  \BibitemOpen
  \bibfield  {author} {\bibinfo {author} {\bibfnamefont {L.}~\bibnamefont
  {Anderegg}}, \bibinfo {author} {\bibfnamefont {L.~W.}\ \bibnamefont {Cheuk}},
  \bibinfo {author} {\bibfnamefont {Y.}~\bibnamefont {Bao}}, \bibinfo {author}
  {\bibfnamefont {S.}~\bibnamefont {Burchesky}}, \bibinfo {author}
  {\bibfnamefont {W.}~\bibnamefont {Ketterle}}, \bibinfo {author}
  {\bibfnamefont {K.~K.}\ \bibnamefont {Ni}},\ and\ \bibinfo {author}
  {\bibfnamefont {J.~M.}\ \bibnamefont {Doyle}},\ }\bibfield  {title} {\bibinfo
  {title} {An optical tweezer array of ultracold molecules},\ }\href@noop {}
  {\bibfield  {journal} {\bibinfo  {journal} {Science}\ }\textbf {\bibinfo
  {volume} {365}},\ \bibinfo {pages} {1156} (\bibinfo {year}
  {2019})}\BibitemShut {NoStop}%
\bibitem [{\citenamefont {Roussy}\ \emph {et~al.}(2022)\citenamefont {Roussy},
  \citenamefont {Caldwell}, \citenamefont {Wright}, \citenamefont {Cairncross},
  \citenamefont {Shagam}, \citenamefont {Ng}, \citenamefont {Scholssberger},
  \citenamefont {Park}, \citenamefont {Wang}, \citenamefont {Ye},\ and\
  \citenamefont {Cornell}}]{rca2022}%
  \BibitemOpen
  \bibfield  {author} {\bibinfo {author} {\bibfnamefont {T.~S.}\ \bibnamefont
  {Roussy}}, \bibinfo {author} {\bibfnamefont {L.}~\bibnamefont {Caldwell}},
  \bibinfo {author} {\bibfnamefont {T.}~\bibnamefont {Wright}}, \bibinfo
  {author} {\bibfnamefont {W.~B.}\ \bibnamefont {Cairncross}}, \bibinfo
  {author} {\bibfnamefont {Y.}~\bibnamefont {Shagam}}, \bibinfo {author}
  {\bibfnamefont {K.~B.}\ \bibnamefont {Ng}}, \bibinfo {author} {\bibfnamefont
  {N.}~\bibnamefont {Scholssberger}}, \bibinfo {author} {\bibfnamefont {S.~Y.}\
  \bibnamefont {Park}}, \bibinfo {author} {\bibfnamefont {A.}~\bibnamefont
  {Wang}}, \bibinfo {author} {\bibfnamefont {J.}~\bibnamefont {Ye}},\ and\
  \bibinfo {author} {\bibfnamefont {E.~A.}\ \bibnamefont {Cornell}},\
  }\bibfield  {title} {\bibinfo {title} {A new bound on the electron's electric
  dipole moment},\ }\href@noop {} {\bibfield  {journal} {\bibinfo  {journal}
  {Science}\ }\textbf {\bibinfo {volume} {381}},\ \bibinfo {pages} {46}
  (\bibinfo {year} {2022})}\BibitemShut {NoStop}%
\bibitem [{\citenamefont {Norrgard}\ \emph {et~al.}(2016)\citenamefont
  {Norrgard}, \citenamefont {McCarron}, \citenamefont {Steinecker},
  \citenamefont {Tarbutt},\ and\ \citenamefont {DeMille}}]{nmd2016}%
  \BibitemOpen
  \bibfield  {author} {\bibinfo {author} {\bibfnamefont {E.~B.}\ \bibnamefont
  {Norrgard}}, \bibinfo {author} {\bibfnamefont {D.~J.}\ \bibnamefont
  {McCarron}}, \bibinfo {author} {\bibfnamefont {M.~H.}\ \bibnamefont
  {Steinecker}}, \bibinfo {author} {\bibfnamefont {M.~R.}\ \bibnamefont
  {Tarbutt}},\ and\ \bibinfo {author} {\bibfnamefont {D.}~\bibnamefont
  {DeMille}},\ }\bibfield  {title} {\bibinfo {title} {Submillikelvin dipolar
  molecules in a radio-frequency magneto-optical trap},\ }\href@noop {}
  {\bibfield  {journal} {\bibinfo  {journal} {Phys. Rev. Lett.}\ }\textbf
  {\bibinfo {volume} {116}},\ \bibinfo {pages} {063004} (\bibinfo {year}
  {2016})}\BibitemShut {NoStop}%
\bibitem [{\citenamefont {Truppe}\ \emph {et~al.}(2017)\citenamefont {Truppe},
  \citenamefont {Williams}, \citenamefont {Hambach}, \citenamefont {Caldwell},
  \citenamefont {Fitch}, \citenamefont {Hinds}, \citenamefont {Sauer},\ and\
  \citenamefont {Tarbutt}}]{twt2017}%
  \BibitemOpen
  \bibfield  {author} {\bibinfo {author} {\bibfnamefont {S.}~\bibnamefont
  {Truppe}}, \bibinfo {author} {\bibfnamefont {H.~J.}\ \bibnamefont
  {Williams}}, \bibinfo {author} {\bibfnamefont {M.}~\bibnamefont {Hambach}},
  \bibinfo {author} {\bibfnamefont {L.}~\bibnamefont {Caldwell}}, \bibinfo
  {author} {\bibfnamefont {N.~J.}\ \bibnamefont {Fitch}}, \bibinfo {author}
  {\bibfnamefont {E.~A.}\ \bibnamefont {Hinds}}, \bibinfo {author}
  {\bibfnamefont {B.~E.}\ \bibnamefont {Sauer}},\ and\ \bibinfo {author}
  {\bibfnamefont {M.~R.}\ \bibnamefont {Tarbutt}},\ }\bibfield  {title}
  {\bibinfo {title} {Molecules cooled below the doppler limit},\ }\href@noop {}
  {\bibfield  {journal} {\bibinfo  {journal} {Nature Phys.}\ }\textbf {\bibinfo
  {volume} {13}},\ \bibinfo {pages} {1173} (\bibinfo {year}
  {2017})}\BibitemShut {NoStop}%
\bibitem [{\citenamefont {Anderegg}\ \emph {et~al.}(2017)\citenamefont
  {Anderegg}, \citenamefont {Augenbraun}, \citenamefont {Chae}, \citenamefont
  {Hemmerling}, \citenamefont {Hutzler}, \citenamefont {Ravi}, \citenamefont
  {Collopy}, \citenamefont {Ye}, \citenamefont {Ketterle},\ and\ \citenamefont
  {Doyle}}]{aad2017}%
  \BibitemOpen
  \bibfield  {author} {\bibinfo {author} {\bibfnamefont {L.}~\bibnamefont
  {Anderegg}}, \bibinfo {author} {\bibfnamefont {B.~L.}\ \bibnamefont
  {Augenbraun}}, \bibinfo {author} {\bibfnamefont {E.}~\bibnamefont {Chae}},
  \bibinfo {author} {\bibfnamefont {B.}~\bibnamefont {Hemmerling}}, \bibinfo
  {author} {\bibfnamefont {N.~R.}\ \bibnamefont {Hutzler}}, \bibinfo {author}
  {\bibfnamefont {A.}~\bibnamefont {Ravi}}, \bibinfo {author} {\bibfnamefont
  {A.}~\bibnamefont {Collopy}}, \bibinfo {author} {\bibfnamefont
  {J.}~\bibnamefont {Ye}}, \bibinfo {author} {\bibfnamefont {W.}~\bibnamefont
  {Ketterle}},\ and\ \bibinfo {author} {\bibfnamefont {J.~M.}\ \bibnamefont
  {Doyle}},\ }\bibfield  {title} {\bibinfo {title} {Radio frequency
  magneto-optical trapping of {CaF} with high density},\ }\href@noop {}
  {\bibfield  {journal} {\bibinfo  {journal} {Phys. Rev. Lett.}\ }\textbf
  {\bibinfo {volume} {119}},\ \bibinfo {pages} {103201} (\bibinfo {year}
  {2017})}\BibitemShut {NoStop}%
\bibitem [{\citenamefont {Ding}\ \emph {et~al.}(2020)\citenamefont {Ding},
  \citenamefont {Wu}, \citenamefont {Finneran}, \citenamefont {Burau},\ and\
  \citenamefont {Ye}}]{dwy2020}%
  \BibitemOpen
  \bibfield  {author} {\bibinfo {author} {\bibfnamefont {S.}~\bibnamefont
  {Ding}}, \bibinfo {author} {\bibfnamefont {Y.}~\bibnamefont {Wu}}, \bibinfo
  {author} {\bibfnamefont {I.~A.}\ \bibnamefont {Finneran}}, \bibinfo {author}
  {\bibfnamefont {J.~J.}\ \bibnamefont {Burau}},\ and\ \bibinfo {author}
  {\bibfnamefont {J.}~\bibnamefont {Ye}},\ }\bibfield  {title} {\bibinfo
  {title} {Sub-doppler cooling and compressed trapping of {YO} molecules at
  $\mu${K} temperatures},\ }\href@noop {} {\bibfield  {journal} {\bibinfo
  {journal} {Phys. Rev. X.}\ }\textbf {\bibinfo {volume} {10}},\ \bibinfo
  {pages} {021049} (\bibinfo {year} {2020})}\BibitemShut {NoStop}%
\bibitem [{\citenamefont {Vilas}\ \emph {et~al.}(2022)\citenamefont {Vilas},
  \citenamefont {Hallas}, \citenamefont {Anderegg}, \citenamefont {Robichaud},
  \citenamefont {Winnicki}, \citenamefont {Mitra},\ and\ \citenamefont
  {Doyle}}]{vhd2022}%
  \BibitemOpen
  \bibfield  {author} {\bibinfo {author} {\bibfnamefont {N.~B.}\ \bibnamefont
  {Vilas}}, \bibinfo {author} {\bibfnamefont {C.}~\bibnamefont {Hallas}},
  \bibinfo {author} {\bibfnamefont {L.}~\bibnamefont {Anderegg}}, \bibinfo
  {author} {\bibfnamefont {P.}~\bibnamefont {Robichaud}}, \bibinfo {author}
  {\bibfnamefont {A.}~\bibnamefont {Winnicki}}, \bibinfo {author}
  {\bibfnamefont {D.}~\bibnamefont {Mitra}},\ and\ \bibinfo {author}
  {\bibfnamefont {J.~M.}\ \bibnamefont {Doyle}},\ }\bibfield  {title} {\bibinfo
  {title} {Magneto-optical trapping and sub-doppler cooling of a polyatomic
  molecule},\ }\href@noop {} {\bibfield  {journal} {\bibinfo  {journal}
  {Nature}\ }\textbf {\bibinfo {volume} {606}},\ \bibinfo {pages} {70}
  (\bibinfo {year} {2022})}\BibitemShut {NoStop}%
\bibitem [{\citenamefont {Caldwell}\ \emph {et~al.}(2019)\citenamefont
  {Caldwell}, \citenamefont {Devlin}, \citenamefont {Williams}, \citenamefont
  {Fitch}, \citenamefont {Hinds}, \citenamefont {Sauer},\ and\ \citenamefont
  {Tarbutt}}]{cdt2019}%
  \BibitemOpen
  \bibfield  {author} {\bibinfo {author} {\bibfnamefont {L.}~\bibnamefont
  {Caldwell}}, \bibinfo {author} {\bibfnamefont {J.}~\bibnamefont {Devlin}},
  \bibinfo {author} {\bibfnamefont {H.}~\bibnamefont {Williams}}, \bibinfo
  {author} {\bibfnamefont {N.}~\bibnamefont {Fitch}}, \bibinfo {author}
  {\bibfnamefont {E.}~\bibnamefont {Hinds}}, \bibinfo {author} {\bibfnamefont
  {B.}~\bibnamefont {Sauer}},\ and\ \bibinfo {author} {\bibfnamefont
  {M.}~\bibnamefont {Tarbutt}},\ }\bibfield  {title} {\bibinfo {title} {Deep
  laser cooling and efficient magnetic compression of molecules},\ }\href@noop
  {} {\bibfield  {journal} {\bibinfo  {journal} {Phys. Rev. Lett.}\ }\textbf
  {\bibinfo {volume} {123}},\ \bibinfo {pages} {033202} (\bibinfo {year}
  {2019})}\BibitemShut {NoStop}%
\bibitem [{\citenamefont {McCarron}\ \emph {et~al.}(2018)\citenamefont
  {McCarron}, \citenamefont {Steinecker}, \citenamefont {Zhu},\ and\
  \citenamefont {DeMille}}]{msd2018}%
  \BibitemOpen
  \bibfield  {author} {\bibinfo {author} {\bibfnamefont {D.~J.}\ \bibnamefont
  {McCarron}}, \bibinfo {author} {\bibfnamefont {M.~H.}\ \bibnamefont
  {Steinecker}}, \bibinfo {author} {\bibfnamefont {Y.}~\bibnamefont {Zhu}},\
  and\ \bibinfo {author} {\bibfnamefont {D.}~\bibnamefont {DeMille}},\
  }\bibfield  {title} {\bibinfo {title} {Magnetic trapping of an ultracold gas
  of polar molecules},\ }\href@noop {} {\bibfield  {journal} {\bibinfo
  {journal} {Phys. Rev. Lett.}\ }\textbf {\bibinfo {volume} {121}},\ \bibinfo
  {pages} {013202} (\bibinfo {year} {2018})}\BibitemShut {NoStop}%
\bibitem [{\citenamefont {Cheuk}\ \emph {et~al.}(2018)\citenamefont {Cheuk},
  \citenamefont {Anderegg}, \citenamefont {Augenbraun}, \citenamefont {Bao},
  \citenamefont {Burchesky}, \citenamefont {Ketterle},\ and\ \citenamefont
  {Doyle}}]{cad2018}%
  \BibitemOpen
  \bibfield  {author} {\bibinfo {author} {\bibfnamefont {L.~W.}\ \bibnamefont
  {Cheuk}}, \bibinfo {author} {\bibfnamefont {L.}~\bibnamefont {Anderegg}},
  \bibinfo {author} {\bibfnamefont {B.~L.}\ \bibnamefont {Augenbraun}},
  \bibinfo {author} {\bibfnamefont {Y.}~\bibnamefont {Bao}}, \bibinfo {author}
  {\bibfnamefont {S.}~\bibnamefont {Burchesky}}, \bibinfo {author}
  {\bibfnamefont {W.}~\bibnamefont {Ketterle}},\ and\ \bibinfo {author}
  {\bibfnamefont {J.~M.}\ \bibnamefont {Doyle}},\ }\bibfield  {title} {\bibinfo
  {title} {{$\Lambda$}-enhanced imaging of molecules in an optical trap},\
  }\href@noop {} {\bibfield  {journal} {\bibinfo  {journal} {Phys. Rev. Lett.}\
  }\textbf {\bibinfo {volume} {121}},\ \bibinfo {pages} {083201} (\bibinfo
  {year} {2018})}\BibitemShut {NoStop}%
\bibitem [{\citenamefont {Wu}\ \emph {et~al.}(2021)\citenamefont {Wu},
  \citenamefont {Burau}, \citenamefont {Mehling}, \citenamefont {Ye},\ and\
  \citenamefont {Ding}}]{wbd2021}%
  \BibitemOpen
  \bibfield  {author} {\bibinfo {author} {\bibfnamefont {Y.}~\bibnamefont
  {Wu}}, \bibinfo {author} {\bibfnamefont {J.~J.}\ \bibnamefont {Burau}},
  \bibinfo {author} {\bibfnamefont {K.}~\bibnamefont {Mehling}}, \bibinfo
  {author} {\bibfnamefont {J.}~\bibnamefont {Ye}},\ and\ \bibinfo {author}
  {\bibfnamefont {S.}~\bibnamefont {Ding}},\ }\bibfield  {title} {\bibinfo
  {title} {High phase-space density of laser-cooled molecules in an optical
  lattice},\ }\href@noop {} {\bibfield  {journal} {\bibinfo  {journal} {Phys.
  Rev. Lett.}\ }\textbf {\bibinfo {volume} {127}},\ \bibinfo {pages} {263201}
  (\bibinfo {year} {2021})}\BibitemShut {NoStop}%
\bibitem [{\citenamefont {Lu}\ \emph {et~al.}(2022)\citenamefont {Lu},
  \citenamefont {Holland},\ and\ \citenamefont {Cheuk}}]{lhc2022}%
  \BibitemOpen
  \bibfield  {author} {\bibinfo {author} {\bibfnamefont {Y.}~\bibnamefont
  {Lu}}, \bibinfo {author} {\bibfnamefont {C.~M.}\ \bibnamefont {Holland}},\
  and\ \bibinfo {author} {\bibfnamefont {L.~W.}\ \bibnamefont {Cheuk}},\
  }\bibfield  {title} {\bibinfo {title} {Molecular laser cooling in a
  dynamically tunable repulsive optical trap},\ }\href@noop {} {\bibfield
  {journal} {\bibinfo  {journal} {Phys. Rev. Lett.}\ }\textbf {\bibinfo
  {volume} {128}},\ \bibinfo {pages} {213201} (\bibinfo {year}
  {2022})}\BibitemShut {NoStop}%
\bibitem [{\citenamefont {Hallas}\ \emph {et~al.}(2023)\citenamefont {Hallas},
  \citenamefont {Vilas}, \citenamefont {Anderegg}, \citenamefont {Robichaud},
  \citenamefont {Winnicki}, \citenamefont {Zhang}, \citenamefont {Cheng},\ and\
  \citenamefont {Doyle}}]{hvd2022}%
  \BibitemOpen
  \bibfield  {author} {\bibinfo {author} {\bibfnamefont {C.}~\bibnamefont
  {Hallas}}, \bibinfo {author} {\bibfnamefont {N.~B.}\ \bibnamefont {Vilas}},
  \bibinfo {author} {\bibfnamefont {L.}~\bibnamefont {Anderegg}}, \bibinfo
  {author} {\bibfnamefont {P.}~\bibnamefont {Robichaud}}, \bibinfo {author}
  {\bibfnamefont {A.}~\bibnamefont {Winnicki}}, \bibinfo {author}
  {\bibfnamefont {C.}~\bibnamefont {Zhang}}, \bibinfo {author} {\bibfnamefont
  {L.}~\bibnamefont {Cheng}},\ and\ \bibinfo {author} {\bibfnamefont {J.~M.}\
  \bibnamefont {Doyle}},\ }\bibfield  {title} {\bibinfo {title} {Optical
  trapping of a polyatomic molecule in an l-type parity doublet state},\
  }\href@noop {} {\bibfield  {journal} {\bibinfo  {journal} {Phys. Rev. Lett.}\
  }\textbf {\bibinfo {volume} {130}},\ \bibinfo {pages} {153202} (\bibinfo
  {year} {2023})}\BibitemShut {NoStop}%
\bibitem [{\citenamefont {Langin}\ \emph {et~al.}(2021)\citenamefont {Langin},
  \citenamefont {Jorapur}, \citenamefont {Zhu}, \citenamefont {Wang},\ and\
  \citenamefont {DeMille}}]{ljd2021}%
  \BibitemOpen
  \bibfield  {author} {\bibinfo {author} {\bibfnamefont {T.~K.}\ \bibnamefont
  {Langin}}, \bibinfo {author} {\bibfnamefont {V.}~\bibnamefont {Jorapur}},
  \bibinfo {author} {\bibfnamefont {Y.}~\bibnamefont {Zhu}}, \bibinfo {author}
  {\bibfnamefont {Q.}~\bibnamefont {Wang}},\ and\ \bibinfo {author}
  {\bibfnamefont {D.}~\bibnamefont {DeMille}},\ }\bibfield  {title} {\bibinfo
  {title} {Polarization enhanced deep optical dipole trapping of
  {$\Lambda-$}cooled polar molecules},\ }\href@noop {} {\bibfield  {journal}
  {\bibinfo  {journal} {Phys. Rev. Lett.}\ }\textbf {\bibinfo {volume} {127}},\
  \bibinfo {pages} {163201} (\bibinfo {year} {2021})}\BibitemShut {NoStop}%
\bibitem [{\citenamefont {Anderson}\ \emph {et~al.}(1995)\citenamefont
  {Anderson}, \citenamefont {Ensher}, \citenamefont {Matthews}, \citenamefont
  {Wieman},\ and\ \citenamefont {Cornell}}]{aec1995}%
  \BibitemOpen
  \bibfield  {author} {\bibinfo {author} {\bibfnamefont {M.~H.}\ \bibnamefont
  {Anderson}}, \bibinfo {author} {\bibfnamefont {J.~R.}\ \bibnamefont
  {Ensher}}, \bibinfo {author} {\bibfnamefont {M.~R.}\ \bibnamefont
  {Matthews}}, \bibinfo {author} {\bibfnamefont {C.~E.}\ \bibnamefont
  {Wieman}},\ and\ \bibinfo {author} {\bibfnamefont {E.~A.}\ \bibnamefont
  {Cornell}},\ }\bibfield  {title} {\bibinfo {title} {Observation of
  bose-einstein condensation in a dilute atomic vapor},\ }\href@noop {}
  {\bibfield  {journal} {\bibinfo  {journal} {Science}\ }\textbf {\bibinfo
  {volume} {269}},\ \bibinfo {pages} {198} (\bibinfo {year}
  {1995})}\BibitemShut {NoStop}%
\bibitem [{\citenamefont {Schindewolf}\ \emph {et~al.}(2022)\citenamefont
  {Schindewolf}, \citenamefont {Bause}, \citenamefont {Chen}, \citenamefont
  {Duda}, \citenamefont {Karman}, \citenamefont {Bloch},\ and\ \citenamefont
  {Luo}}]{sbl2022}%
  \BibitemOpen
  \bibfield  {author} {\bibinfo {author} {\bibfnamefont {A.}~\bibnamefont
  {Schindewolf}}, \bibinfo {author} {\bibfnamefont {R.}~\bibnamefont {Bause}},
  \bibinfo {author} {\bibfnamefont {X.}~\bibnamefont {Chen}}, \bibinfo {author}
  {\bibfnamefont {M.}~\bibnamefont {Duda}}, \bibinfo {author} {\bibfnamefont
  {T.}~\bibnamefont {Karman}}, \bibinfo {author} {\bibfnamefont
  {I.}~\bibnamefont {Bloch}},\ and\ \bibinfo {author} {\bibfnamefont
  {X.}~\bibnamefont {Luo}},\ }\bibfield  {title} {\bibinfo {title} {Evaporation
  of microwave-shielded polar molecules to quantum degeneracy},\ }\href@noop {}
  {\bibfield  {journal} {\bibinfo  {journal} {Nature}\ }\textbf {\bibinfo
  {volume} {607}},\ \bibinfo {pages} {677} (\bibinfo {year}
  {2022})}\BibitemShut {NoStop}%
\bibitem [{\citenamefont {Gregory}\ \emph {et~al.}(2019)\citenamefont
  {Gregory}, \citenamefont {Frye}, \citenamefont {Blackmore}, \citenamefont
  {Bridge}, \citenamefont {Sawant}, \citenamefont {Hutson},\ and\ \citenamefont
  {Cornish}}]{gfc2019}%
  \BibitemOpen
  \bibfield  {author} {\bibinfo {author} {\bibfnamefont {P.~D.}\ \bibnamefont
  {Gregory}}, \bibinfo {author} {\bibfnamefont {M.~D.}\ \bibnamefont {Frye}},
  \bibinfo {author} {\bibfnamefont {J.~A.}\ \bibnamefont {Blackmore}}, \bibinfo
  {author} {\bibfnamefont {E.~M.}\ \bibnamefont {Bridge}}, \bibinfo {author}
  {\bibfnamefont {R.}~\bibnamefont {Sawant}}, \bibinfo {author} {\bibfnamefont
  {J.~M.}\ \bibnamefont {Hutson}},\ and\ \bibinfo {author} {\bibfnamefont
  {S.~L.}\ \bibnamefont {Cornish}},\ }\bibfield  {title} {\bibinfo {title}
  {Sticky collisions of ultracold {RbCs} molecules},\ }\href@noop {} {\bibfield
   {journal} {\bibinfo  {journal} {Nat Commun}\ }\textbf {\bibinfo {volume}
  {10}},\ \bibinfo {pages} {3104} (\bibinfo {year} {2019})}\BibitemShut
  {NoStop}%
\bibitem [{\citenamefont {He}\ \emph {et~al.}(2021)\citenamefont {He},
  \citenamefont {Ye}, \citenamefont {Lin}, \citenamefont {Guo}, \citenamefont
  {Quéméner},\ and\ \citenamefont {Wang}}]{hyw2021}%
  \BibitemOpen
  \bibfield  {author} {\bibinfo {author} {\bibfnamefont {J.}~\bibnamefont
  {He}}, \bibinfo {author} {\bibfnamefont {X.}~\bibnamefont {Ye}}, \bibinfo
  {author} {\bibfnamefont {J.}~\bibnamefont {Lin}}, \bibinfo {author}
  {\bibfnamefont {M.}~\bibnamefont {Guo}}, \bibinfo {author} {\bibfnamefont
  {G.}~\bibnamefont {Quéméner}},\ and\ \bibinfo {author} {\bibfnamefont
  {D.}~\bibnamefont {Wang}},\ }\bibfield  {title} {\bibinfo {title}
  {Observation of resonant dipolar collisions in ultracold
  $^{23}${Na}$^{87}${Rb} rotational mixtures},\ }\href@noop {} {\bibfield
  {journal} {\bibinfo  {journal} {Phys. Rev. Research}\ }\textbf {\bibinfo
  {volume} {3}},\ \bibinfo {pages} {013016} (\bibinfo {year}
  {2021})}\BibitemShut {NoStop}%
\bibitem [{\citenamefont {Voges}\ \emph {et~al.}(2020)\citenamefont {Voges},
  \citenamefont {Gersema}, \citenamefont {zum Alten~Borgloh}, \citenamefont
  {Schulze}, \citenamefont {Hartmann}, \citenamefont {Zenesini},\ and\
  \citenamefont {Ospelkaus}}]{vgo2020}%
  \BibitemOpen
  \bibfield  {author} {\bibinfo {author} {\bibfnamefont {K.~K.}\ \bibnamefont
  {Voges}}, \bibinfo {author} {\bibfnamefont {P.}~\bibnamefont {Gersema}},
  \bibinfo {author} {\bibfnamefont {M.~M.}\ \bibnamefont {zum Alten~Borgloh}},
  \bibinfo {author} {\bibfnamefont {T.~A.}\ \bibnamefont {Schulze}}, \bibinfo
  {author} {\bibfnamefont {T.}~\bibnamefont {Hartmann}}, \bibinfo {author}
  {\bibfnamefont {A.}~\bibnamefont {Zenesini}},\ and\ \bibinfo {author}
  {\bibfnamefont {S.}~\bibnamefont {Ospelkaus}},\ }\bibfield  {title} {\bibinfo
  {title} {Ultracold gas of bosonic $^{23}${Na}$^{39}${K} ground-state
  molecules},\ }\href@noop {} {\bibfield  {journal} {\bibinfo  {journal} {Phys.
  Rev. Lett.}\ }\textbf {\bibinfo {volume} {125}},\ \bibinfo {pages} {083401}
  (\bibinfo {year} {2020})}\BibitemShut {NoStop}%
\bibitem [{\citenamefont {Bause}\ \emph {et~al.}(2023)\citenamefont {Bause},
  \citenamefont {Christianen}, \citenamefont {Schindewolf}, \citenamefont
  {Bloch},\ and\ \citenamefont {Luo}}]{bcl2023}%
  \BibitemOpen
  \bibfield  {author} {\bibinfo {author} {\bibfnamefont {R.}~\bibnamefont
  {Bause}}, \bibinfo {author} {\bibfnamefont {A.}~\bibnamefont {Christianen}},
  \bibinfo {author} {\bibfnamefont {A.}~\bibnamefont {Schindewolf}}, \bibinfo
  {author} {\bibfnamefont {I.}~\bibnamefont {Bloch}},\ and\ \bibinfo {author}
  {\bibfnamefont {X.}~\bibnamefont {Luo}},\ }\bibfield  {title} {\bibinfo
  {title} {Ultracold sticky collisions: Theoretical and experimental status},\
  }\href@noop {} {\bibfield  {journal} {\bibinfo  {journal} {J. Phys. Chem.}\
  }\textbf {\bibinfo {volume} {127}},\ \bibinfo {pages} {729} (\bibinfo {year}
  {2023})}\BibitemShut {NoStop}%
\bibitem [{\citenamefont {Segev}\ \emph {et~al.}(2019)\citenamefont {Segev},
  \citenamefont {Pitzer}, \citenamefont {Karpov}, \citenamefont {Akerman},
  \citenamefont {Narevicius},\ and\ \citenamefont {Narevicius}}]{spn2019}%
  \BibitemOpen
  \bibfield  {author} {\bibinfo {author} {\bibfnamefont {Y.}~\bibnamefont
  {Segev}}, \bibinfo {author} {\bibfnamefont {M.}~\bibnamefont {Pitzer}},
  \bibinfo {author} {\bibfnamefont {M.}~\bibnamefont {Karpov}}, \bibinfo
  {author} {\bibfnamefont {N.}~\bibnamefont {Akerman}}, \bibinfo {author}
  {\bibfnamefont {J.}~\bibnamefont {Narevicius}},\ and\ \bibinfo {author}
  {\bibfnamefont {E.}~\bibnamefont {Narevicius}},\ }\bibfield  {title}
  {\bibinfo {title} {Collisions between cold molecules in a superconducting
  magnetic trap},\ }\href@noop {} {\bibfield  {journal} {\bibinfo  {journal}
  {Nature}\ }\textbf {\bibinfo {volume} {572}},\ \bibinfo {pages} {189}
  (\bibinfo {year} {2019})}\BibitemShut {NoStop}%
\bibitem [{\citenamefont {Cheuk}\ \emph {et~al.}(2020)\citenamefont {Cheuk},
  \citenamefont {Anderegg}, \citenamefont {Bao}, \citenamefont {Burchesky},
  \citenamefont {Yu}, \citenamefont {Ketterle}, \citenamefont {Ni},\ and\
  \citenamefont {Doyle}}]{cad2020}%
  \BibitemOpen
  \bibfield  {author} {\bibinfo {author} {\bibfnamefont {L.~W.}\ \bibnamefont
  {Cheuk}}, \bibinfo {author} {\bibfnamefont {L.}~\bibnamefont {Anderegg}},
  \bibinfo {author} {\bibfnamefont {Y.}~\bibnamefont {Bao}}, \bibinfo {author}
  {\bibfnamefont {S.}~\bibnamefont {Burchesky}}, \bibinfo {author}
  {\bibfnamefont {S.}~\bibnamefont {Yu}}, \bibinfo {author} {\bibfnamefont
  {W.}~\bibnamefont {Ketterle}}, \bibinfo {author} {\bibfnamefont {K.~K.}\
  \bibnamefont {Ni}},\ and\ \bibinfo {author} {\bibfnamefont {J.~M.}\
  \bibnamefont {Doyle}},\ }\bibfield  {title} {\bibinfo {title} {Observation of
  collisions between two ultracold ground-state {CaF} molecules},\ }\href@noop
  {} {\bibfield  {journal} {\bibinfo  {journal} {Phys. Rev. Lett.}\ }\textbf
  {\bibinfo {volume} {125}},\ \bibinfo {pages} {043401} (\bibinfo {year}
  {2020})}\BibitemShut {NoStop}%
\bibitem [{\citenamefont {Takekoshi}\ \emph {et~al.}(2014)\citenamefont
  {Takekoshi}, \citenamefont {Reichsöllner}, \citenamefont {Schindewolf},
  \citenamefont {Hutson}, \citenamefont {Sueur}, \citenamefont {Dulieu},
  \citenamefont {Ferlaino}, \citenamefont {Grimm},\ and\ \citenamefont
  {Nägerl}}]{trn2014}%
  \BibitemOpen
  \bibfield  {author} {\bibinfo {author} {\bibfnamefont {T.}~\bibnamefont
  {Takekoshi}}, \bibinfo {author} {\bibfnamefont {L.}~\bibnamefont
  {Reichsöllner}}, \bibinfo {author} {\bibfnamefont {A.}~\bibnamefont
  {Schindewolf}}, \bibinfo {author} {\bibfnamefont {J.~M.}\ \bibnamefont
  {Hutson}}, \bibinfo {author} {\bibfnamefont {C.~R.~L.}\ \bibnamefont
  {Sueur}}, \bibinfo {author} {\bibfnamefont {O.}~\bibnamefont {Dulieu}},
  \bibinfo {author} {\bibfnamefont {F.}~\bibnamefont {Ferlaino}}, \bibinfo
  {author} {\bibfnamefont {R.}~\bibnamefont {Grimm}},\ and\ \bibinfo {author}
  {\bibfnamefont {H.~C.}\ \bibnamefont {Nägerl}},\ }\bibfield  {title}
  {\bibinfo {title} {Ultracold dense samples of dipolar {RbCs} molecules in the
  rovibrational and hyperfine ground state},\ }\href@noop {} {\bibfield
  {journal} {\bibinfo  {journal} {Phys. Rev. Lett.}\ }\textbf {\bibinfo
  {volume} {113}},\ \bibinfo {pages} {205301} (\bibinfo {year}
  {2014})}\BibitemShut {NoStop}%
\bibitem [{\citenamefont {Park}\ \emph {et~al.}(2015)\citenamefont {Park},
  \citenamefont {Will},\ and\ \citenamefont {Zwierlein}}]{paz2015}%
  \BibitemOpen
  \bibfield  {author} {\bibinfo {author} {\bibfnamefont {J.~W.}\ \bibnamefont
  {Park}}, \bibinfo {author} {\bibfnamefont {S.~A.}\ \bibnamefont {Will}},\
  and\ \bibinfo {author} {\bibfnamefont {M.~W.}\ \bibnamefont {Zwierlein}},\
  }\bibfield  {title} {\bibinfo {title} {Ultracold dipolar gas of fermionic
  $^{23}${Na}$^{40}${K} molecules in their absolute ground state},\ }\href@noop
  {} {\bibfield  {journal} {\bibinfo  {journal} {Phys. Rev. Lett.}\ }\textbf
  {\bibinfo {volume} {114}},\ \bibinfo {pages} {205302} (\bibinfo {year}
  {2015})}\BibitemShut {NoStop}%
\bibitem [{\citenamefont {Karman}\ and\ \citenamefont
  {Hutson}(2018)}]{khu2018}%
  \BibitemOpen
  \bibfield  {author} {\bibinfo {author} {\bibfnamefont {T.}~\bibnamefont
  {Karman}}\ and\ \bibinfo {author} {\bibfnamefont {J.~M.}\ \bibnamefont
  {Hutson}},\ }\bibfield  {title} {\bibinfo {title} {Microwave shielding of
  ultracold polar molecules},\ }\href@noop {} {\bibfield  {journal} {\bibinfo
  {journal} {Phys. Rev. Lett.}\ }\textbf {\bibinfo {volume} {121}},\ \bibinfo
  {pages} {163401} (\bibinfo {year} {2018})}\BibitemShut {NoStop}%
\bibitem [{\citenamefont {Lassabliere}\ and\ \citenamefont
  {Quemener}(2018)}]{lqu2018}%
  \BibitemOpen
  \bibfield  {author} {\bibinfo {author} {\bibfnamefont {L.}~\bibnamefont
  {Lassabliere}}\ and\ \bibinfo {author} {\bibfnamefont {G.}~\bibnamefont
  {Quemener}},\ }\bibfield  {title} {\bibinfo {title} {Controlling the
  scattering length of ultracold dipolar molecules},\ }\href@noop {} {\bibfield
   {journal} {\bibinfo  {journal} {Phys. Rev. Lett.}\ }\textbf {\bibinfo
  {volume} {121}},\ \bibinfo {pages} {163402} (\bibinfo {year}
  {2018})}\BibitemShut {NoStop}%
\bibitem [{\citenamefont {Anderegg}\ \emph {et~al.}(2021)\citenamefont
  {Anderegg}, \citenamefont {Burchesky}, \citenamefont {Bao}, \citenamefont
  {Yu}, \citenamefont {Karman}, \citenamefont {Chae}, \citenamefont {Ni},
  \citenamefont {Ketterle},\ and\ \citenamefont {Doyle}}]{abd2021}%
  \BibitemOpen
  \bibfield  {author} {\bibinfo {author} {\bibfnamefont {L.}~\bibnamefont
  {Anderegg}}, \bibinfo {author} {\bibfnamefont {S.}~\bibnamefont {Burchesky}},
  \bibinfo {author} {\bibfnamefont {Y.}~\bibnamefont {Bao}}, \bibinfo {author}
  {\bibfnamefont {S.}~\bibnamefont {Yu}}, \bibinfo {author} {\bibfnamefont
  {T.}~\bibnamefont {Karman}}, \bibinfo {author} {\bibfnamefont
  {E.}~\bibnamefont {Chae}}, \bibinfo {author} {\bibfnamefont {K.~K.}\
  \bibnamefont {Ni}}, \bibinfo {author} {\bibfnamefont {W.}~\bibnamefont
  {Ketterle}},\ and\ \bibinfo {author} {\bibfnamefont {J.~M.}\ \bibnamefont
  {Doyle}},\ }\bibfield  {title} {\bibinfo {title} {Observation of microwave
  shielding of ultracold molecules},\ }\href@noop {} {\bibfield  {journal}
  {\bibinfo  {journal} {Science}\ }\textbf {\bibinfo {volume} {373}},\ \bibinfo
  {pages} {779} (\bibinfo {year} {2021})}\BibitemShut {NoStop}%
\bibitem [{\citenamefont {Lin}\ \emph {et~al.}(2023)\citenamefont {Lin},
  \citenamefont {Chen}, \citenamefont {Jin}, \citenamefont {Shi}, \citenamefont
  {Deng}, \citenamefont {Zhang}, \citenamefont {Quéméner}, \citenamefont
  {Shi}, \citenamefont {Yi},\ and\ \citenamefont {Wang}}]{lcw2023}%
  \BibitemOpen
  \bibfield  {author} {\bibinfo {author} {\bibfnamefont {J.}~\bibnamefont
  {Lin}}, \bibinfo {author} {\bibfnamefont {G.}~\bibnamefont {Chen}}, \bibinfo
  {author} {\bibfnamefont {M.}~\bibnamefont {Jin}}, \bibinfo {author}
  {\bibfnamefont {Z.}~\bibnamefont {Shi}}, \bibinfo {author} {\bibfnamefont
  {F.}~\bibnamefont {Deng}}, \bibinfo {author} {\bibfnamefont {W.}~\bibnamefont
  {Zhang}}, \bibinfo {author} {\bibfnamefont {G.}~\bibnamefont {Quéméner}},
  \bibinfo {author} {\bibfnamefont {T.}~\bibnamefont {Shi}}, \bibinfo {author}
  {\bibfnamefont {S.}~\bibnamefont {Yi}},\ and\ \bibinfo {author}
  {\bibfnamefont {D.}~\bibnamefont {Wang}},\ }\bibfield  {title} {\bibinfo
  {title} {Microwave shielding of bosonic {NaRb} molecules},\ }\href@noop {}
  {\bibfield  {journal} {\bibinfo  {journal} {arXiv:2304.08312v2}\ } (\bibinfo
  {year} {2023})}\BibitemShut {NoStop}%
\bibitem [{\citenamefont {Bigagli}\ \emph {et~al.}(2023)\citenamefont
  {Bigagli}, \citenamefont {Warner}, \citenamefont {Yuan}, \citenamefont
  {Zhang}, \citenamefont {Stevenson}, \citenamefont {Karman},\ and\
  \citenamefont {Will}}]{bww2023}%
  \BibitemOpen
  \bibfield  {author} {\bibinfo {author} {\bibfnamefont {N.}~\bibnamefont
  {Bigagli}}, \bibinfo {author} {\bibfnamefont {C.}~\bibnamefont {Warner}},
  \bibinfo {author} {\bibfnamefont {W.}~\bibnamefont {Yuan}}, \bibinfo {author}
  {\bibfnamefont {S.}~\bibnamefont {Zhang}}, \bibinfo {author} {\bibfnamefont
  {I.}~\bibnamefont {Stevenson}}, \bibinfo {author} {\bibfnamefont
  {T.}~\bibnamefont {Karman}},\ and\ \bibinfo {author} {\bibfnamefont
  {S.}~\bibnamefont {Will}},\ }\bibfield  {title} {\bibinfo {title}
  {Collisionally stable gas of bosonic dipolar ground state molecules},\
  }\href@noop {} {\bibfield  {journal} {\bibinfo  {journal}
  {arXiv:2303.16845v1}\ } (\bibinfo {year} {2023})}\BibitemShut {NoStop}%
\bibitem [{\citenamefont {Quéméner}\ and\ \citenamefont
  {Bohn}(2016)}]{gb2016}%
  \BibitemOpen
  \bibfield  {author} {\bibinfo {author} {\bibfnamefont {G.}~\bibnamefont
  {Quéméner}}\ and\ \bibinfo {author} {\bibfnamefont {J.~L.}\ \bibnamefont
  {Bohn}},\ }\bibfield  {title} {\bibinfo {title} {Shielding $^2{\Sigma}$
  ultracold dipolar molecular collisions with electric fields},\ }\href@noop {}
  {\bibfield  {journal} {\bibinfo  {journal} {Phys. Rev. A.}\ }\textbf
  {\bibinfo {volume} {93}},\ \bibinfo {pages} {012704} (\bibinfo {year}
  {2016})}\BibitemShut {NoStop}%
\bibitem [{\citenamefont {Li}\ \emph {et~al.}(2021)\citenamefont {Li},
  \citenamefont {Tobias}, \citenamefont {Matsuda}, \citenamefont {Miller},
  \citenamefont {Valotlina}, \citenamefont {Marco}, \citenamefont {Wang},
  \citenamefont {Lassabliere}, \citenamefont {Quemener}, \citenamefont {Bohn},\
  and\ \citenamefont {Ye}}]{lty2021}%
  \BibitemOpen
  \bibfield  {author} {\bibinfo {author} {\bibfnamefont {J.}~\bibnamefont
  {Li}}, \bibinfo {author} {\bibfnamefont {W.~G.}\ \bibnamefont {Tobias}},
  \bibinfo {author} {\bibfnamefont {K.}~\bibnamefont {Matsuda}}, \bibinfo
  {author} {\bibfnamefont {C.}~\bibnamefont {Miller}}, \bibinfo {author}
  {\bibfnamefont {G.}~\bibnamefont {Valotlina}}, \bibinfo {author}
  {\bibfnamefont {L.~D.}\ \bibnamefont {Marco}}, \bibinfo {author}
  {\bibfnamefont {R.}~\bibnamefont {Wang}}, \bibinfo {author} {\bibfnamefont
  {L.}~\bibnamefont {Lassabliere}}, \bibinfo {author} {\bibfnamefont
  {G.}~\bibnamefont {Quemener}}, \bibinfo {author} {\bibfnamefont {J.~L.}\
  \bibnamefont {Bohn}},\ and\ \bibinfo {author} {\bibfnamefont
  {J.}~\bibnamefont {Ye}},\ }\bibfield  {title} {\bibinfo {title} {Tuning of
  dipolar interactions and evaporative cooling in a three-dimensional molecular
  quantum gas},\ }\href@noop {} {\bibfield  {journal} {\bibinfo  {journal}
  {Nature Phys.}\ }\textbf {\bibinfo {volume} {17}},\ \bibinfo {pages} {1144}
  (\bibinfo {year} {2021})}\BibitemShut {NoStop}%
\bibitem [{\citenamefont {Jurgilas}\ \emph {et~al.}(2021)\citenamefont
  {Jurgilas}, \citenamefont {Chakraborty}, \citenamefont {Rich}, \citenamefont
  {Caldwell}, \citenamefont {Williams}, \citenamefont {Fitch}, \citenamefont
  {Sauer}, \citenamefont {Frye}, \citenamefont {Hutson},\ and\ \citenamefont
  {Tarbutt}}]{jct2021}%
  \BibitemOpen
  \bibfield  {author} {\bibinfo {author} {\bibfnamefont {S.}~\bibnamefont
  {Jurgilas}}, \bibinfo {author} {\bibfnamefont {A.}~\bibnamefont
  {Chakraborty}}, \bibinfo {author} {\bibfnamefont {C.}~\bibnamefont {Rich}},
  \bibinfo {author} {\bibfnamefont {L.}~\bibnamefont {Caldwell}}, \bibinfo
  {author} {\bibfnamefont {H.}~\bibnamefont {Williams}}, \bibinfo {author}
  {\bibfnamefont {N.}~\bibnamefont {Fitch}}, \bibinfo {author} {\bibfnamefont
  {B.}~\bibnamefont {Sauer}}, \bibinfo {author} {\bibfnamefont {M.~D.}\
  \bibnamefont {Frye}}, \bibinfo {author} {\bibfnamefont {J.~M.}\ \bibnamefont
  {Hutson}},\ and\ \bibinfo {author} {\bibfnamefont {M.}~\bibnamefont
  {Tarbutt}},\ }\bibfield  {title} {\bibinfo {title} {Collisions between
  ultracold molecules and atoms in a magnetic trap},\ }\href@noop {} {\bibfield
   {journal} {\bibinfo  {journal} {Phys. Rev. Lett.}\ }\textbf {\bibinfo
  {volume} {126}},\ \bibinfo {pages} {153401} (\bibinfo {year}
  {2021})}\BibitemShut {NoStop}%
\bibitem [{\citenamefont {Devlin}\ and\ \citenamefont
  {Tarbutt}(2016)}]{dta2016}%
  \BibitemOpen
  \bibfield  {author} {\bibinfo {author} {\bibfnamefont {J.~A.}\ \bibnamefont
  {Devlin}}\ and\ \bibinfo {author} {\bibfnamefont {M.~R.}\ \bibnamefont
  {Tarbutt}},\ }\bibfield  {title} {\bibinfo {title} {Three-dimensional
  doppler, polarization-gradient, and magneto-optical forces for atoms and
  molecules with dark states},\ }\href@noop {} {\bibfield  {journal} {\bibinfo
  {journal} {New J. Phys.}\ }\textbf {\bibinfo {volume} {18}},\ \bibinfo
  {pages} {123017} (\bibinfo {year} {2016})}\BibitemShut {NoStop}%
\bibitem [{\citenamefont {Devlin}\ and\ \citenamefont
  {Tarbutt}(2018)}]{dta2018}%
  \BibitemOpen
  \bibfield  {author} {\bibinfo {author} {\bibfnamefont {J.~A.}\ \bibnamefont
  {Devlin}}\ and\ \bibinfo {author} {\bibfnamefont {M.~R.}\ \bibnamefont
  {Tarbutt}},\ }\bibfield  {title} {\bibinfo {title} {Laser cooling and
  magneto-optical trapping of molecules analyzed using optical bloch equations
  and the fokker-planck-kramers equation},\ }\href@noop {} {\bibfield
  {journal} {\bibinfo  {journal} {Phys. Rev. A.}\ }\textbf {\bibinfo {volume}
  {98}},\ \bibinfo {pages} {063415} (\bibinfo {year} {2018})}\BibitemShut
  {NoStop}%
\bibitem [{\citenamefont {Langin}\ and\ \citenamefont
  {DeMille}(2023)}]{lde2023}%
  \BibitemOpen
  \bibfield  {author} {\bibinfo {author} {\bibfnamefont {T.~K.}\ \bibnamefont
  {Langin}}\ and\ \bibinfo {author} {\bibfnamefont {D.}~\bibnamefont
  {DeMille}},\ }\bibfield  {title} {\bibinfo {title} {Toward improved loading,
  cooling, and trapping of molecules in magneto-optical traps},\ }\href@noop {}
  {\bibfield  {journal} {\bibinfo  {journal} {New J. Phys.}\ }\textbf {\bibinfo
  {volume} {25}},\ \bibinfo {pages} {043005} (\bibinfo {year}
  {2023})}\BibitemShut {NoStop}%
\bibitem [{\citenamefont {Callopy}\ \emph {et~al.}(2018)\citenamefont
  {Callopy}, \citenamefont {Ding}, \citenamefont {Wu}, \citenamefont
  {Finneran}, \citenamefont {Anderegg}, \citenamefont {Augenbraun},
  \citenamefont {Doyle},\ and\ \citenamefont {Ye}}]{cdy2018}%
  \BibitemOpen
  \bibfield  {author} {\bibinfo {author} {\bibfnamefont {A.~L.}\ \bibnamefont
  {Callopy}}, \bibinfo {author} {\bibfnamefont {S.}~\bibnamefont {Ding}},
  \bibinfo {author} {\bibfnamefont {Y.}~\bibnamefont {Wu}}, \bibinfo {author}
  {\bibfnamefont {I.~A.}\ \bibnamefont {Finneran}}, \bibinfo {author}
  {\bibfnamefont {L.}~\bibnamefont {Anderegg}}, \bibinfo {author}
  {\bibfnamefont {B.~L.}\ \bibnamefont {Augenbraun}}, \bibinfo {author}
  {\bibfnamefont {J.~M.}\ \bibnamefont {Doyle}},\ and\ \bibinfo {author}
  {\bibfnamefont {J.}~\bibnamefont {Ye}},\ }\bibfield  {title} {\bibinfo
  {title} {{3D} magneto-optical trap of yttrium monoxide},\ }\href@noop {}
  {\bibfield  {journal} {\bibinfo  {journal} {Phys. Rev. Lett.}\ }\textbf
  {\bibinfo {volume} {121}},\ \bibinfo {pages} {213201} (\bibinfo {year}
  {2018})}\BibitemShut {NoStop}%
\bibitem [{\citenamefont {Jarvis}\ \emph {et~al.}(2018)\citenamefont {Jarvis},
  \citenamefont {Devlin}, \citenamefont {Wall}, \citenamefont {Sauer},\ and\
  \citenamefont {Tarbutt}}]{jdt2018}%
  \BibitemOpen
  \bibfield  {author} {\bibinfo {author} {\bibfnamefont {K.~N.}\ \bibnamefont
  {Jarvis}}, \bibinfo {author} {\bibfnamefont {J.~A.}\ \bibnamefont {Devlin}},
  \bibinfo {author} {\bibfnamefont {T.~E.}\ \bibnamefont {Wall}}, \bibinfo
  {author} {\bibfnamefont {B.~E.}\ \bibnamefont {Sauer}},\ and\ \bibinfo
  {author} {\bibfnamefont {M.~R.}\ \bibnamefont {Tarbutt}},\ }\bibfield
  {title} {\bibinfo {title} {Blue-detuned magneto-optical trap},\ }\href@noop
  {} {\bibfield  {journal} {\bibinfo  {journal} {Phys. Rev. Lett.}\ }\textbf
  {\bibinfo {volume} {120}},\ \bibinfo {pages} {083201} (\bibinfo {year}
  {2018})}\BibitemShut {NoStop}%
\bibitem [{\citenamefont {Burau}\ \emph {et~al.}(2023)\citenamefont {Burau},
  \citenamefont {Aggarwal}, \citenamefont {Mehling},\ and\ \citenamefont
  {Ye}}]{bay2023}%
  \BibitemOpen
  \bibfield  {author} {\bibinfo {author} {\bibfnamefont {J.~J.}\ \bibnamefont
  {Burau}}, \bibinfo {author} {\bibfnamefont {P.}~\bibnamefont {Aggarwal}},
  \bibinfo {author} {\bibfnamefont {K.}~\bibnamefont {Mehling}},\ and\ \bibinfo
  {author} {\bibfnamefont {J.}~\bibnamefont {Ye}},\ }\bibfield  {title}
  {\bibinfo {title} {Blue-detuned magneto-optical trap of molecules},\
  }\href@noop {} {\bibfield  {journal} {\bibinfo  {journal} {Phys. Rev. Lett.}\
  }\textbf {\bibinfo {volume} {130}},\ \bibinfo {pages} {193401} (\bibinfo
  {year} {2023})}\BibitemShut {NoStop}%
\bibitem [{\citenamefont {Barry}\ \emph {et~al.}(2014)\citenamefont {Barry},
  \citenamefont {McCarron}, \citenamefont {Norrgard}, \citenamefont
  {Steinecker},\ and\ \citenamefont {DeMille}}]{bmd2014}%
  \BibitemOpen
  \bibfield  {author} {\bibinfo {author} {\bibfnamefont {J.~F.}\ \bibnamefont
  {Barry}}, \bibinfo {author} {\bibfnamefont {D.~J.}\ \bibnamefont {McCarron}},
  \bibinfo {author} {\bibfnamefont {E.~B.}\ \bibnamefont {Norrgard}}, \bibinfo
  {author} {\bibfnamefont {M.~H.}\ \bibnamefont {Steinecker}},\ and\ \bibinfo
  {author} {\bibfnamefont {D.}~\bibnamefont {DeMille}},\ }\bibfield  {title}
  {\bibinfo {title} {Magneto-optical trapping of a diatomic molecule},\
  }\href@noop {} {\bibfield  {journal} {\bibinfo  {journal} {Nature}\ }\textbf
  {\bibinfo {volume} {512}},\ \bibinfo {pages} {286} (\bibinfo {year}
  {2014})}\BibitemShut {NoStop}%
\bibitem [{\citenamefont {Barry}\ \emph {et~al.}(2012)\citenamefont {Barry},
  \citenamefont {Shuman}, \citenamefont {Norrgard},\ and\ \citenamefont
  {DeMille}}]{bsd2012}%
  \BibitemOpen
  \bibfield  {author} {\bibinfo {author} {\bibfnamefont {J.~F.}\ \bibnamefont
  {Barry}}, \bibinfo {author} {\bibfnamefont {E.~S.}\ \bibnamefont {Shuman}},
  \bibinfo {author} {\bibfnamefont {E.~B.}\ \bibnamefont {Norrgard}},\ and\
  \bibinfo {author} {\bibfnamefont {D.}~\bibnamefont {DeMille}},\ }\bibfield
  {title} {\bibinfo {title} {Laser radiation pressure slowing of a molecular
  beam},\ }\href@noop {} {\bibfield  {journal} {\bibinfo  {journal} {Phys. Rev.
  Lett.}\ }\textbf {\bibinfo {volume} {108}},\ \bibinfo {pages} {103002}
  (\bibinfo {year} {2012})}\BibitemShut {NoStop}%
\bibitem [{\citenamefont {Tarbutt}\ and\ \citenamefont
  {Steimle}(2015)}]{tst2015}%
  \BibitemOpen
  \bibfield  {author} {\bibinfo {author} {\bibfnamefont {M.~R.}\ \bibnamefont
  {Tarbutt}}\ and\ \bibinfo {author} {\bibfnamefont {T.~C.}\ \bibnamefont
  {Steimle}},\ }\bibfield  {title} {\bibinfo {title} {Modeling magneto-optical
  trapping of caf molecules},\ }\href@noop {} {\bibfield  {journal} {\bibinfo
  {journal} {Phys. Rev. A.}\ }\textbf {\bibinfo {volume} {92}},\ \bibinfo
  {pages} {053401} (\bibinfo {year} {2015})}\BibitemShut {NoStop}%
\bibitem [{\citenamefont {Barry}(2013)}]{barry}%
  \BibitemOpen
  \bibfield  {author} {\bibinfo {author} {\bibfnamefont {J.~F.}\ \bibnamefont
  {Barry}},\ }\emph {\bibinfo {title} {Laser cooling and slowing of a diatomic
  molecule}},\ \href@noop {} {Ph.D. thesis},\ \bibinfo  {school} {Yale
  University}, \bibinfo {address} {New Haven CT} (\bibinfo {year}
  {2013})\BibitemShut {NoStop}%
\bibitem [{sup()}]{supp}%
  \BibitemOpen
  \bibfield  {title} {\bibinfo {title} {See supplemental material at [url will
  be inserted by publisher] for information on the blue-mot, the data taking
  procedure and the calculation of the unitarity limit},\ }\href@noop {} {\
  }\BibitemShut {NoStop}%
\bibitem [{\citenamefont {{NL-eEDM Collaboration}}(2019)}]{NLEDM2019}%
  \BibitemOpen
  \bibfield  {author} {\bibinfo {author} {\bibnamefont {{NL-eEDM
  Collaboration}}},\ }\bibfield  {title} {\bibinfo {title} {High accuracy
  theoretical investigations of {CaF}, {SrF}, and {BaF} and implications for
  laser-cooling},\ }\href@noop {} {\bibfield  {journal} {\bibinfo  {journal}
  {J. Chem. Phys.}\ }\textbf {\bibinfo {volume} {151}},\ \bibinfo {pages}
  {034302} (\bibinfo {year} {2019})}\BibitemShut {NoStop}%
\bibitem [{\citenamefont {Meyer}\ and\ \citenamefont {Bohn}(2011)}]{mb2011}%
  \BibitemOpen
  \bibfield  {author} {\bibinfo {author} {\bibfnamefont {E.~R.}\ \bibnamefont
  {Meyer}}\ and\ \bibinfo {author} {\bibfnamefont {J.~L.}\ \bibnamefont
  {Bohn}},\ }\bibfield  {title} {\bibinfo {title} {Chemical pathways in
  ultracold reactions of {SrF} molecules},\ }\href@noop {} {\bibfield
  {journal} {\bibinfo  {journal} {Phys. Rev. A.}\ }\textbf {\bibinfo {volume}
  {83}},\ \bibinfo {pages} {032714} (\bibinfo {year} {2011})}\BibitemShut
  {NoStop}%
\bibitem [{\citenamefont {Lepers}\ and\ \citenamefont {Dulieu}(2017)}]{ld2017}%
  \BibitemOpen
  \bibfield  {author} {\bibinfo {author} {\bibfnamefont {M.}~\bibnamefont
  {Lepers}}\ and\ \bibinfo {author} {\bibfnamefont {O.}~\bibnamefont
  {Dulieu}},\ }\bibfield  {title} {\bibinfo {title} {Long-range interactions
  between ultracold atoms and molecules},\ }\href@noop {} {\bibfield  {journal}
  {\bibinfo  {journal} {arXiv:1703.02833}\ } (\bibinfo {year}
  {2017})}\BibitemShut {NoStop}%
\bibitem [{\citenamefont {Quéméner}\ \emph {et~al.}(2011)\citenamefont
  {Quéméner}, \citenamefont {Bohn}, \citenamefont {Petrov},\ and\
  \citenamefont {Kotochigova}}]{qbpk2011}%
  \BibitemOpen
  \bibfield  {author} {\bibinfo {author} {\bibfnamefont {G.}~\bibnamefont
  {Quéméner}}, \bibinfo {author} {\bibfnamefont {J.~L.}\ \bibnamefont
  {Bohn}}, \bibinfo {author} {\bibfnamefont {A.}~\bibnamefont {Petrov}},\ and\
  \bibinfo {author} {\bibfnamefont {S.}~\bibnamefont {Kotochigova}},\
  }\bibfield  {title} {\bibinfo {title} {Universalities in ultracold reactions
  of alkali-metal polar molecules},\ }\href@noop {} {\bibfield  {journal}
  {\bibinfo  {journal} {Phys. Rev. A.}\ }\textbf {\bibinfo {volume} {84}},\
  \bibinfo {pages} {062703} (\bibinfo {year} {2011})}\BibitemShut {NoStop}%
\end{thebibliography}%

\end{document}